%% file: target_selection.tex
\documentclass[usenatbib,useAMS]{mn2e}
\usepackage[titletoc,toc]{appendix}
\usepackage{epsfig,psfig,graphicx,float,color}
\usepackage{subfigure}
\usepackage{mybibdefs}
\usepackage{multirow}
\usepackage{widetext}
\usepackage{url}
\usepackage{color}
\usepackage{amsmath}

\newcommand\rr{\color{black}}
\newcommand\rrr{\color{black}}

\usepackage{graphicx}
\usepackage{amssymb}
\usepackage{epstopdf}
\title[Machine learning target selection]{Tuning target selection algorithms to improve galaxy redshift estimates}  
\author[Hoyle et al.]{Ben  Hoyle$^{1,2}$, Kerstin Paech$^{1,2}$,  Markus Michael Rau$^{1,3}$,   \newauthor Stella Seitz$^{1,3}$, Jochen Weller$^{1,2,3}$\\\\
$^1$Universitaets-Sternwarte, Fakultaet fuer Physik, Ludwig-Maximilians Universitaet Muenchen, Scheinerstr. 1, D-81679 Muenchen, Germany\\
$^2$Excellence Cluster Universe, Boltzmannstr. 2, D-85748 Garching, Germany\\
$^3$Max Planck Institute fuer Extraterrestrial Physics, Giessenbachstr. 1, D-85748 Garching, Germany\\
\\
{\tt E-mail: hoyleb@usm.uni-muenchen.de}
} 
  
\begin{document}
\date{Accepted ----. Received ----; in original form ----.}
\pagerange{\pageref{firstpage}--\pageref{lastpage}} \pubyear{2010}
\maketitle
\label{firstpage}
\begin{abstract}
We showcase machine learning (ML) inspired target selection algorithms to determine which of all potential targets should be selected first for spectroscopic follow up.  Efficient target selection can improve the ML redshift uncertainties as calculated on an independent sample, while requiring less targets to be observed.  We compare  {\rr 7 different} ML targeting algorithms with the Sloan Digital Sky Survey (SDSS) target order, and with a random targeting algorithm. The ML inspired algorithms are constructed iteratively by estimating which of the remaining target galaxies will be most difficult for the  machine learning methods to accurately estimate redshifts using the previously observed data. This is performed by predicting the expected redshift error and redshift offset (or bias) of all of the remaining target galaxies. We find that the predicted values of bias and error are accurate to better than 10-30\% of the true values, even with only limited training sample sizes.  We construct a hypothetical follow-up survey and find that some of the ML targeting algorithms are able to obtain the same redshift predictive power with 2-3 times less observing time, as compared to that of the SDSS, or random, target selection algorithms. The reduction in the required follow up resources could allow for a change to the follow-up strategy, for example by obtaining deeper spectroscopy, which could improve ML redshift estimates for deeper test data.

\end{abstract}
\begin{keywords}
galaxies: distances and redshifts,  catalogues, surveys.
\end{keywords}

\section{introduction}
\input{intro.tex}

\section{Data and Machine learning methods}
\label{data}
\input{data.tex}
\section{Method and Analysis}
\label{method}
\input{method.tex}
\section{Results}
\label{results}
\input{results.tex}
\section{Summary \& conclusions}
\label{conclusions}
\input{conclus.tex}

\section*{Acknowledgments} 
\label{ack}
The authors would like to thank an anonymous referee for comments and suggestions which have improve the paper. B. Hoyle, S.Seitz and M. M. Rau acknowledge support from the Transregional Collaborative
Research Centre TRR 33 - The Dark Universe and the DFG
cluster of excellence ``Origin and Structure of the Universe''. 
Funding for the SDSS and SDSS-II has been provided by the Alfred
P. Sloan Foundation, the Participating Institutions, the
National Science Foundation, the U.S. Department of
Energy, the National Aeronautics and Space Administration,
the Japanese Monbukagakusho, the Max Planck
Society, and the Higher Education Funding Council for
England. The SDSS Web Site is http://www.sdss.org/. 
\bibliographystyle{mn2e}
\bibliography{photoz}

\end{document}

%% file: intro.tex
In order to maximise the cosmological information from large scale structure surveys, samples of galaxies must be identified and their positions on the sky, photometric properties, and redshifts measured. Measuring accurate spectroscopic redshifts is resource intensive and is typically only performed for a small sub sample of all galaxies. For this sub sample of galaxies one may learn a mapping between the measured photometric properties (or `features'), and the spectroscopic redshift. The learnt mapping can then be applied to all photometrically identified galaxies to estimate photometric redshifts.  

This paper aims to address the problem of identifying which galaxies, from an available target list, should be targeted first for spectroscopic follow up, in order to reduce the uncertainty on the machine learning redshift estimates of a final test sample. To examine this problem, and to test our methods, we construct sets of simulated (or hypothetical) observing runs using existing data.

Depending on the science case, one may wish to target particular types of galaxies for spectroscopic follow-up. The authors \cite{2014MNRAS.438.2218J} use artificial Neural Networks in a classification analysis to select potential targets using photometric data, that are consistent with being emission line galaxies.

Within the science case of attempting to improve the photometric redshifts of sub samples of galaxies, \citep{tpz} suggest a method to use random forests to estimate photometric redshift pdfs. The authors identify volumes in color-magnitude space which produce poor redshift estimates, as defined by the compactness of the photometric redshift probability distribution function (hereafter pdf). They suggest that targeting galaxies in these color cells can lead to improvement in the photometric redshift of other galaxies within the cell. The authors \citep{2015ApJ...813...53M} use a similar technique by compressing a high dimensional color space into complex two dimensional cells using Self Organised Maps \citep{Kohonen:1997:SM:261082}. The authors determine which regions of the projected feature space have cells which are populated by a test sample galaxy, but are not currently populated by a spectroscopic training example. Underpopulated cells can then be targeted for follow up, however determining the density of spectra in a cell is difficult because of the non trivial cell volume. The method presented in this paper differs from these approaches. We attempt to predict which galaxies should be targeted next, and in which order, from a list of all possible targets, such that the redshift metrics on a final test sample are iteratively improved. This differs from the above approaches by not requiring that spectra are taken in certain feature cells, if they are not required to improve redshift estimates, or that some cells could still be targeted further if this helps the remaining sample to have improved redshift estimates. We also investigate different targeting algorithms and compare their performance using standard redshift performance metrics.

Machine learning methods and techniques have long been applied to photometric redshifts analysis since \cite{2003LNCS.2859..226T} used artificial Neural Networks. Since this time a plethora of machine learning architectures, including the tree based methods used in this work, have been 
applied to the problem of point prediction redshift estimation or to estimate the redshift pdf \citep[see e.g.,][]{2004PASP..116..345C,2007AN....328..852C,2008ASPC..394..521C,2010ApJ...715..823G,2013arXiv1312.1287B,tpz,2015MNRAS.449.2040H,RauEtAllinPrep}. Further concepts  have also recently been ported from machine learning to astronomy, such as feature importance \citep{2014arXiv1410.4696H}, feature extraction \citep{2014ASPC..485..425P}, data augmentation \citep{2004A&A...423..761V,2015arXiv150106759H} and deep machine learning \citep{2014arXiv1412.8341H,2015arXiv150307077D,2015arXiv150407255H}. Machine learning architectures have also had success in other fields of astronomy such as galaxy morphology identification, and star \& quasar separation \citep[see for example][]{1997daa..conf...43L,2009arXiv0910.3770Y}. Other data driven, albeit non machine learning, redshift estimates techniques exist and are gaining popularity \citep[e.g., clustering techniques][]{2013arXiv1303.4722M}.

The original target selection algorithm for the Sloan Digital Sky Survey \citep[hereafter SDSS,][]{2000AJ....120.1579Y} was proposed to fulfill a broad range of criteria, including;  the uniformity of the sample, the insensitivity to systematic effects such as seeing,  the requirement that it should be based on physically meaningful parameters which correlated with galaxy properties, and that it should be as simple as possible to allow for the construction of mock catalogues \citep{2002AJ....124.1810S}. Afterwards other targeting algorithms were suggested to construct samples of Luminous Red Galaxies \citep{2001AJ....122.2267E} and Quasars \citep{2002AJ....123.2945R}. This paper uses SDSS data to  explore other targeting algorithms, which are tuned to reduce the redshift uncertainty of an independent test sample of representative galaxies.

This paper is organized as follows: In \S\ref{data} we describe the data sample and the machine learning methods employed, the design of the experiments and their analysis in \S\ref{method}, and results in \S\ref{results}. We summarise and conclude  in \S\ref{conclusions}.

%% file: data.tex
In this study we use observational data drawn from the SDSS III Data Release 12 \citep[][]{2015arXiv150100963A}, and divide the sample into two sub groups by the date when the spectra were taken. These sub samples correspond to approximately SDSS I\&II and SDSS III. We analyse each sub sample separately during independent sets of analyses.

\subsection{Observational dataset}
\label{obs_data}
The SDSS I-III uses a {\rr 2.5} meter telescope at Apache Point Observatory in New Mexico and has CCD, wide field photometry, in 5 bands \citep[$u,g,r,i,z$][]{Gunn:2006tw,Smith:2002pca}, and an expansive spectroscopic follow up program \citep[][]{2011AJ....142...72E} covering $\pi$ steradians of the northern and equatorial sky. The SDSS collaboration has obtained more than 3 million galaxy spectra using dual fiber-fed spectrographs. An automated photometric pipeline performs object classification to a magnitude of $r\approx22$ and measures photometric properties of more than 100 million galaxies. The complete data sample, and many derived catalogs such as galaxy photometric properties, are publicly available through the {\tt CasJobs} server \citep[][]{10.1109/MCSE.2008.6}\footnote{skyserver.sdss3.org/CasJobs}.

The SDSS dataset is well suited for the analyses presented in this paper due to the large number of photometrically selected galaxies with spectroscopic redshifts to use as training and test samples and the documented date corresponding to when the spectra were taken.  These spectral dates are used as one of the comparison target selection algorithms, described further in \S\ref{method}.

We select all objects from {\tt CasJobs} with both spectroscopic redshifts and photometric properties. In detail we run the {\tt MySQL} query shown in the appendix resulting in 3,751,496 objects. We next select all of the 2,183,897 objects which are classified by the photometric pipeline {\tt PHOTPTYPE} as being a galaxy, and have measured errors on the model magnitudes in all $g,r,i$ bands to be less than 0.2. We next select all of the 2,158,880 objects with a spectroscopic redshift error below 0.001, and with spectroscopic redshift to be above 0, and to have no spectral warning flags set. Finally we remove duplicate photometric {\tt objid} which reduces the sample to 1,981,397. This selects a clean sample of galaxies, free of stellar contamination, which is suitable for the analysis in this work. {\rr One} may also choose to select combinations of stars, galaxies, and other artefacts if the science goal were to improve star/galaxy separation. {\rr We note that stellar contamination, or contamination by AGN is a very real problem which plagues target selection routines and photometric redshift estimates, especially at fainter magnitudes. However while non galaxy sources may contaminate the galaxy sample in reality, they are important to observe, to construct a representative samples in order to accurately perform star, AGN and galaxy separation, in combination with accurately estimating redshifts.}

From this base set we construct two samples, based on the MJD time stamp that the galaxy spectra was taken. We  select 869,479 objects to approximately construct the SDSS I\&II sample by selecting spectra with MJD values to be less than 5.47e4, and 1,111,918 spectra with MJD values greater than 5.47e4 to be the SDSS III sample.  We randomly extract 100,000 galaxies from both samples to form the list of potential training set target galaxies, and use the remaining galaxies as the final test sets to estimate redshift prediction ability in each set of analyses. We have also explored using 80\% of the galaxy sample as the potential target galaxies and find very similar results. However using these larger samples leads to a dramatic increase in computation time due to the number of machine learning systems we construct for each of the hypothetical observing runs, as described in \S\ref{treemethods} and \S\ref{method}.

In this work we have concentrated on the following six typical features for redshift estimation; the $r$ band magnitude, the following colours: $r-g$, $g-u$, $i-r$, $z-r$, and the Petrosian radius measured in the $r$ band. Previous work has shown that there are many other readily obtained photometric features which also have strong predictive power when estimating redshifts \citep[][]{2014arXiv1410.4696H}, however defining an optimal redshift estimation technique for the data sample in this paper is not the main focus of this {\rr work}.
 
\subsection{Tree based methods}
\label{treemethods}
We use the scikit-learn implementation of decision trees for regression \citep[][]{ig} as the machine learning architecture to predict galaxy redshifts, redshift errors, and redshift biases. The decision tree algorithm recursively partitions the input feature dimensions into an increasing number of bins. {\rr The bin boundaries are chosen to minimize the scatter of the output feature for all of the object which sit in each bin. In this work the output feature} can be the spectroscopic redshift, the photometric redshift bias, or the photometric redshift error (see \S\ref{method}). 

The power of tree based methods is enhanced by combining many trees. One technique to do this is called Random Forests \citep[][]{RandoMforests} which constructs $N$ trees simultaneously by drawing random selections of both feature space and random (with replacement) samples of training data $D_{tr}$, with which to train each tree. Each tree is grown by recursively partitioning the selected property (or `feature') with the choice of partition centre being performed using a `greedy' strategy, in order to reduce the variance of the output feature between the data in each partition. The data which sit on each final leaf node form the prediction value for new data $D$, which is queried down the tree. {\rr  Each tree $T$, can be viewed as learning a model of the data, but is prone to either over fitting, or not modelling the complexity in the data well enough. However, the combined prediction from many trees produces a forest prediction $P$ which is a model with strong predictive power, and is not prone to over fitting. The random forest prediction is obtained for new data $D$  following
}

\begin{equation}
P(D) = \frac{1}{N}\sum_{i=0}^{N} T_i(D).
\label{equ:forest}
\end{equation}
 In a regression analysis, such as that used to predict redshift estimates, the final output value is therefore the average value from each tree in the forest.  We also measure the standard deviation of predictions from all of the trees for each galaxy $g$, and refer to this quantity as the photometric error, defined as $\delta_{z_{ML}}=\sigma(T(D_g))$. 

 For more details about constructing random forests we refer the reader to \cite{hastie01statisticallearning}\footnote{\url{statweb.stanford.edu/~tibs/ElemStatLearn}}. The hyper-parameters of the random forest include the final number of data on each leaf node $n_L$, the number of trees in the forest $N$, and the maximum number of input features that are separately selected during the training of each tree in the forest, $n_F$.

In all of the work which follows, and when referring to machine learning systems, we will imply the use of random forests. For each random forest system, we actually construct eight random forests, each with a random choice of hyper-parameter values, and we select the best fitting random forest for the task at hand by using a cross validation hold-out sample. We do not perform a grid search of hyper-parameter space due to the large number of systems which are trained. {\rr We normally find that after training 30 machines with randomly selected hyper-parameters, the best trained machine produces metric values which are typically very close to that of a comparable system, but with more than 100 random hyper-parameter selections. Our choice of 8 random samples of hyper-parameter ensures that we still explore the space well, but not exhaustively, given the number of systems that need to be train. We also note that the hyper-parameters of the best fitting model is always non trivial to predict a priori.} We define the best random forest to be the one with the smallest value of the harmonic mean of the 68\% dispersion (denoted $\sigma_{68}$) and the 95\% dispersion ($\sigma_{95}$) of the measured quantity. The harmonic mean is defined by $\sigma_{68}\times\sigma_{95}/(\sigma_{68}+\sigma_{95})$. We choose to randomly explore the following forest hyper-parameters during this process; the number of trees in the forest from $30<N<250$, the minimum number $n_L$ of data on each leaf node of the tree to be between $1<n_L<100$, and the maximum number of features $n_F\leq 6$ which may be chosen during the random feature selection.

%% file: method.tex
In this section we define and motivate the different target selection algorithms used in this work. In \S \ref{poorTargets} we build a system to predict which targets will be the most difficult to accurately estimate a redshift, and quantify this prediction process in \S\ref{quantpred}. We describe how the different target selection algorithms use these predictions to identify samples for follow up in \S\ref{selection}. We finally describe how the simulated observing runs are performed in \S\ref{setup}.

\subsection{How to identify poor targets}
\label{poorTargets}
To identify which targets will be the most difficult to accurately estimate a redshift we use machine learning systems in two distinct stages, which we outline here, and describe in more detail in \S\ref{prediction}. 

This method is somewhat different from a standard machine learning approach, in which one attempts to estimate the photometric redshift $z_{ML}$, of a galaxy, and to estimate the photometric redshift error $\delta_{z_{ML}}$. Here we rather attempt to predict how large the photometric redshift error $\delta_{z_{ML}}$ on each galaxy will be, and also to predict the offset, or bias, between the redshift point prediction and the true spectroscopic redshift, defined as $b_{z_{ML}} = |z_{spec}-z_{ML}|$.

To achieve these aims, random forests are first used in a standard redshift estimation process using 85\% of the training sample, which iteratively grows after each simulated observing run. The remaining 15\% of the training data is then passed through this first stage and photometric redshifts are estimated. We then calculate the true redshift offset, and redshift error using the 15\% training sample. We train the next machine learning system to predict these values of photometric redshift errors and offsets of this 15\% training sample using only the photometric input features. 

All of the galaxies in the potential target list are passed through this second system to predict and identify which targets will have large redshift errors, and large offsets. Many of these poorly performing targets are selected by the targeting algorithms and then `observed' to obtain a spectroscopic redshift. A schematic diagram of this process is shown in Fig. \ref{MLflowChart}. 

Finally a third machine learning system  is trained for redshift prediction using all of the training galaxies, including the newly observed targets. The photometric properties of the test sample are passed into this third system and redshifts estimates are obtained. We use the test sample redshift predictions to measure performance metrics, see \S\ref{metrics} for details.

We do not attempt to predict the full shape of the redshift probability distribution function for each target because it would induce unnecessary uncertainty. By concentrating on the descriptive point estimates $b_{z_{ML}},\delta_{z_{ML}}$, the signal to noise of the predictions are increased. Calculating these point predictions is also less computationally intensive, and requires less storage space, while also removing the need to select an appropriate band width smoothing scale required for a probability distribution function \citep[see, e.g.][]{RauEtAllinPrep}.

\subsubsection{Predicting poorly estimated targets}
\label{prediction}
Each set of predictions is remade before each simulated observing run and draws from spectra which have been observed up to, and including, the most recent run. Two training samples of size 85\% and 15\% are randomly constructed from this base sample.  {\rr The 85\%/15\% sample split choice is rather arbitrary, but does ensure that most of the data is used to estimate accurate redshifts. The 15\% sample is still of a reasonable size and quickly grows to be more than a few thousand objects, after just over 10\% of the simulated follow up program.} The first 85\% of the training data is used to train a machine to estimate a completely standard photometric redshift $z_{ML}$,  and photometric redshift error $\delta_{z_{ML}}$ using the input photometric features $\Theta$, and the spectroscopic redshift $z_{spec}$, as shown in the upper rectangular box in the schematic diagram in Fig. \ref{MLflowChart}. The 15\% held out sample is passed through the learnt redshift machine to predict both a photometric redshift $z_{ML}$, and error $\delta_{z_{ML}}$. The redshift offset $b_{z_{ML}} = |z_{spec}-z_{ML}|$ of this 15\% sample is calculated exactly, by using the spectroscopic redshifts. {\rr A second round of machine learning systems are next trained to learn the mapping between the input photometric features $\Theta$, and both $b_{z_{ML}}$ and $\delta_{z_{ML}}$ separately. For succinctness, this has been shown by the single lower rectangular box in Fig. \ref{MLflowChart}.}

We subsequently pass the photometric features $\Theta$ of all potential target candidates, which have not yet been observed in our experiment, through the second machine learning system to estimate the redshift offset $b^{est}_{z_{ML}}$, and the redshift error $\delta^{est}_{z_{ML}}$, see the bottom of Fig. \ref{MLflowChart}. We will next use these predictions in \S \ref{selection} to select the poorest performing potential targets using a variety of metrics. But first, we would like to present the following short interlude. Because this is a controlled experiment, and we know the true redshift of all of the targets, we can examine how well we are able to predict the two quantities: redshift offset $b^{est}_{z_{ML}}$, and the redshift error $\delta^{est}_{z_{ML}}$, of the potential target sample.

\subsubsection{Quantifying the prediction performance}
\label{quantpred}
The comparison between the predicted values of offset and bias, and the true values, is performed by first passing the target's photometric quantities $\Theta$ through the first system to measure the photometric redshift $z_{ML}$ and the `true' redshift error $\delta_{z_{ML}}$. The `true' offset is then constructed using $b_{z_{ML}}= |z_{spec}-z_{ML}|$. The photometric quantities $\Theta$ of the targets are next passed through the second system to estimate the offset $b^{est}_{z_{ML}}$, and error  $\delta^{est}_{z_{ML}}$. 

Residual vectors are then constructed by subtracting the estimated value from the `true' value, for all targets, e.g., 
\begin{equation}
\bar{b} = \big| b^{est}_{z_{ML}} - b_{z_{ML}}\big | \; \;, \; \;\bar{\delta} = \big(\delta^{est}_{z_{ML}} - \delta_{z_{ML}}\big). 
\label{equ:delta_b}
\end{equation}
\begin{figure}
 \includegraphics[scale=0.47,clip=true,trim=50 115 50 50]{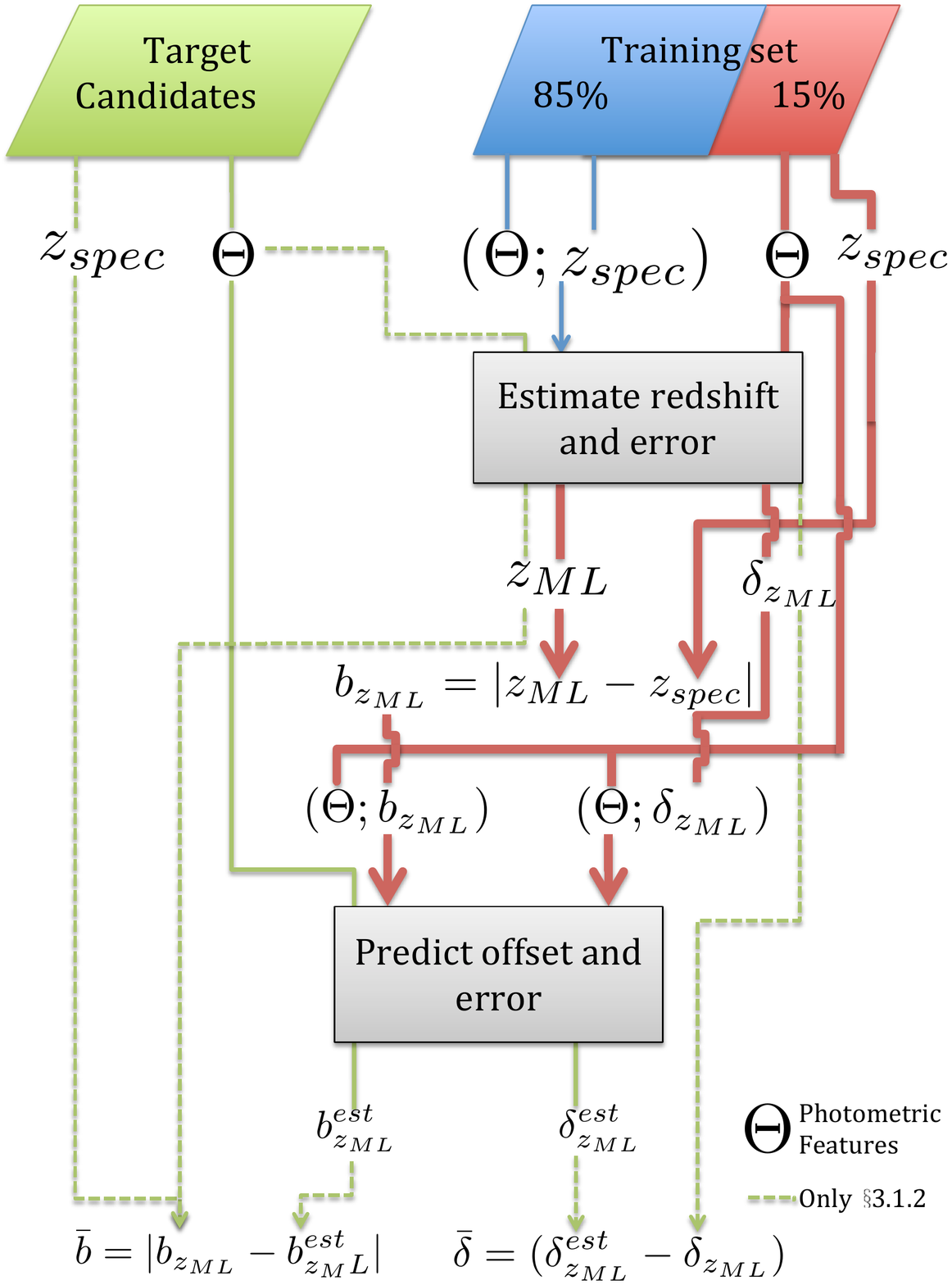}
  \caption{ \label{MLflowChart} A schematic diagram illustrating the prediction of the photometric redshift offset $b^{est}_{z_{ML}}$, and the photometric redshift error $\delta^{est}_{z_{ML}}$, constructed using the training data. The photometric properties of the target candidates are passed though the systems to estimate offsets and errors. Only in \S\ref{quantpred} are these estimates compared with the true values using $\bar{b}$ and $\bar{\delta}$, for this controlled experiment. {\rr We refer the reader to \S\ref{poorTargets} for a fuller description.}}
\end{figure}
These steps are shown on the left hand side, and by the green lines emanating from the target candidates data set, {\rr on} the schematic diagram in Fig. \ref{MLflowChart}. We reiterate that the targets are not used during the training of either of the two machine learning stages.

\begin{figure}
 \includegraphics[scale=0.43,clip=true,trim=0 15 35 30]{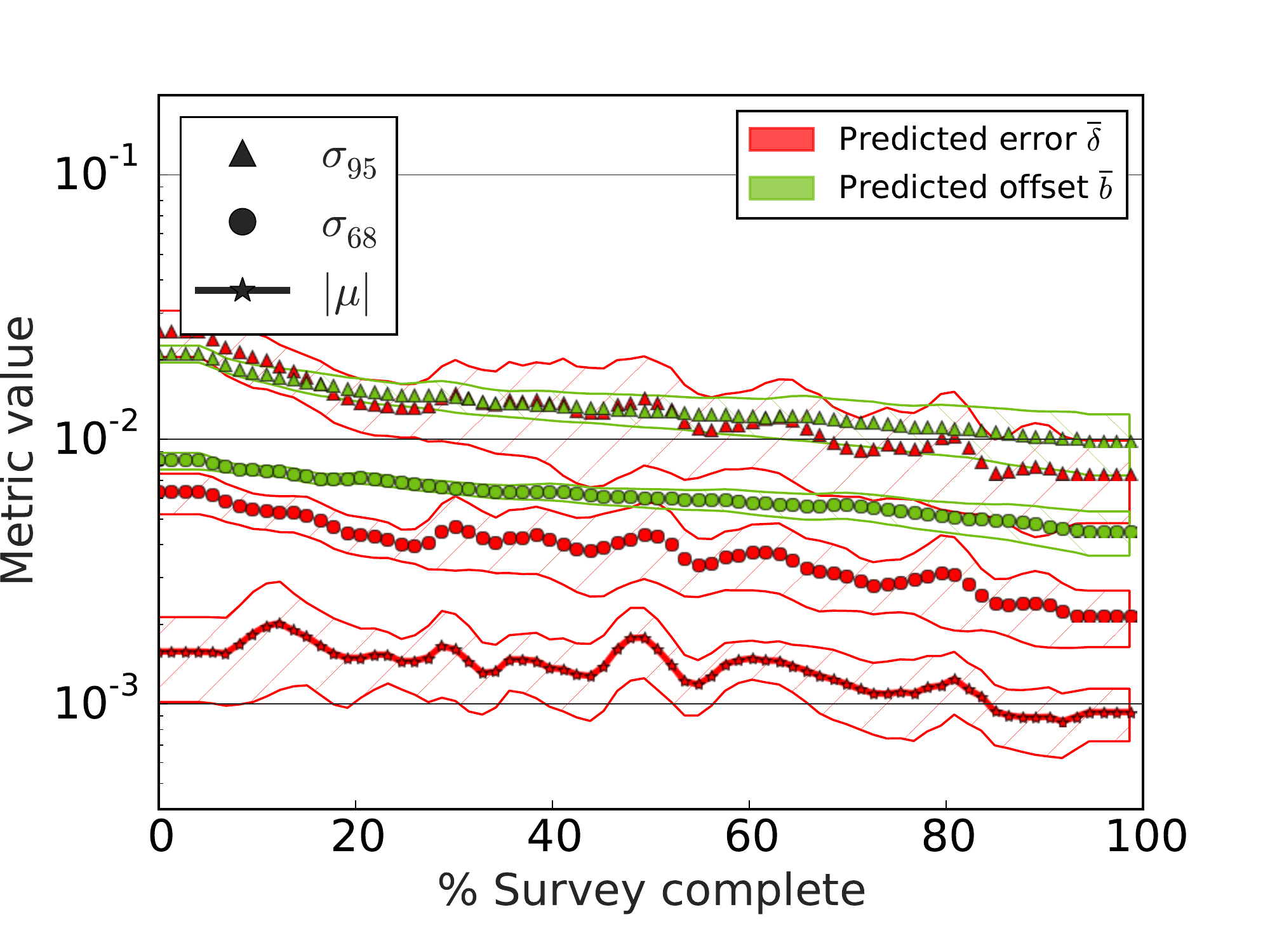}
  \caption{ \label{idError} The ability of the machine learning systems to predict the photometric redshift offset $b^{est}_{z_{ML}}$, and the photometric redshift error $\delta^{est}_{z_{ML}}$, of the SDSS I\&II galaxies which are not used during training. The metrics (see legend) are measured on the residual vectors $\bar{b}$ and $\bar{\delta}$, which are defined in the text. The x-axis describes how many sets of simulated observing runs have been completed in relation to the lifetime of the survey. The hashed contours correspond to the 68\% spread of each of these statistics as measured by the {\rr 7 machine learning targeting} algorithms.}
\end{figure}

To quantify the predictive power of this process on the SDSS I\&II data we measure the metrics $\sigma_{68}$, $\sigma_{95}$, {\rr on the residual vectors $\bar{b}$ and $\bar{\delta}$, and also the median\footnote{ Note that the median value of the predicted offsets $\bar{b}$, corresponds to the Median Absolute Deviation, or MAD, because of the absolute value in the definition.} of the distribution of  $\bar{\delta}$, denoted by $\mu$,}  and present them in Fig. \ref{idError}. The x-axis of Fig. \ref{idError} describes how many sets of simulated observing runs have been completed in relation to the lifetime of the survey.  The data points correspond to the mean of the measured quantities and the hashed contours correspond to the 68\% spread of each of the metric values as measured by {\rr the 7 different machine learning targeting} algorithms, which are described in \S\ref{selection}. 

Fig. \ref{idError} shows that after each simulated observing run the ability of the systems to estimate the redshift offset $b^{est}_{z_{ML}}$, and the redshift error $\delta^{est}_{z_{ML}}$, incrementally improves. The lighter green symbols and hashed regions in Fig. \ref{idError} show that random forests are able to predict the value of the measured redshift offset to within $9\times 10^{-3}$ in 68\% of cases, as shown by $\sigma_{68}$ in the figure, and requires less than 13k galaxies (or 10\% of the total number of simulated observing runs) to achieve this accuracy. Furthermore the 95\% spread of the predicted photometric offset as shown by $\sigma_{95}$, is within 0.04. If we compare these values with typical photometric redshift predictions which have an accuracy of between 0.02 and 0.05 (for the SDSS sample), we find an accuracy within about 30\% of the true value. The darker red symbols and hashed regions present the prediction of the redshift error. We find that the median value of the redshift error, drops below $2\times 10^{-3}$ with even limited training data, and the 68\% spread of predictions (as labelled by $\sigma_{68}$),  is below $5\times 10^{-3}$ within less than 15\% of total observing runs, or 20k galaxies. This is very encouraging and suggests that machine learning is able to correctly estimate the size of the photometric redshift error of a representative, but unseen galaxy, to within $(2\pm5)\times 10^{-3}$, which is over an order of magnitude smaller than the true redshift error value. We note that the prediction ability depends on the choice of targeting selection algorithm because each one will follow up different galaxies during different observing runs, thus compiling a different training sample. The spread of the hashed regions show what effect this has on the measured metrics between the different targeting algorithms. We note that the SDSS III predictions are qualitatively very similar to those of SDSS I\&II as shown in Fig. \ref{idError}, but are approximately a factor of two times higher in each metric value. The flatness of all of the lines at both sides of the figure is an artefact of the smoothing criteria chosen.

\subsection{Target selection algorithms}
\label{selection}
In \S\ref{prediction} we estimate the offset $b^{est}_{z_{ML}}$ and the redshift error $\delta^{est}_{z_{ML}}$ for the remaining potential targets. We now construct different selection algorithms to determine the optimal selection order for the subsequent simulated observing runs. We explore the following set of algorithms to determine which galaxies will be subsequently targeted, and summarise these selection algorithms in Table \ref{TargetSelectionCheatSheet}.

1) We choose N target galaxies in the date order that they were observed by the SDSS.
\\This represents one of the comparison methods. Again we note that the SDSS targeting algorithm which we are replicating in this method was not optimised for machine learning redshift estimation.  

2) We choose target galaxies by randomly selecting N times without replacement from all remaining potential targets.
\\This strategy represents the second comparison method. It addresses the question; how well would we have performed, if we had selected targets at random from the list of potential candidates?

3) We choose target galaxies by selecting the N galaxies with the largest estimated value of redshift offset $b^{est}_{z_{ML}}$. 
\\We could expect this strategy to reduce the outlier fraction of a test sample of galaxies, because these cases will have high values of redshift offset.

4) We choose target galaxies by selecting the N galaxies with the largest estimated value of redshift error $\delta^{est}_{z_{ML}}$.
\\This targeting strategy selects targets with the largest estimated redshift error. We could expect this strategy to reduce the value of the dispersion $\sigma_{68}$ or $\sigma_{95}$ of a test sample of galaxies.

5) We select target galaxies by selecting the N galaxies with the largest estimated value of the t-statistic $T_s$, which is defined as the ratio of offset to error, $T_s$=$b^{est}_{z_{ML}}$/$\delta^{est}_{z_{ML}}$.
\\This targeting strategy examines the cases for which the estimated redshift error is in the least agreement with the estimated redshift offset. This algorithm can be viewed as selecting the extreme tails of the t-distribution. 

6) We target galaxies by selecting the N galaxies with the largest estimated value of the harmonic mean $H_m$ between offset and error, which is defined as, $H_m$= ($b^{est}_{z_{ML}}\times \delta^{est}_{z_{ML}}$)/($b^{est}_{z_{ML}}+\delta^{est}_{z_{ML}}$).
\\This targeting strategy gives equal weight to selecting galaxies with large values of both estimated redshift offset and estimated redshift error. 

7) We select target galaxies by randomly drawing N galaxies from the binned distribution of the t-statistic $T_s$, with the probability of being selected proportional to the inverse of the number of data in the $T_s$ bin.
\\This targeting strategy begins by binning targets in the t-statistic values, and selects targets from the full distribution. However the random selection is chosen to give more weight to the outlying bins because these bins have the fewest number of objects. This statistic is chosen to determine if one can reduce both the outlier fraction and the $\sigma_{68},\sigma_{95}$ dispersions simultaneously.

8) We choose target galaxies by randomly drawing N galaxies from the binned distribution of the harmonic mean $H_m$ between offset and error, with the probability of being selected again proportional to the inverse of the number of data in the $H_m$ bin.
\\This targeting strategy again begins by binning targets in the values of the harmonic mean, and selects targets from the full distribution. The random selection is again chosen to give more weight to the outlying bins because these bins also have the fewest number of objects. This statistic is also chosen to determine if one can reduce both the outlier fraction and the $\sigma_{68},\sigma_{95}$ dispersions simultaneously.

9) We finally also generate a hybrid method which combines equal measures of randomly selected targets and targets with large estimated redshift errors.
\\This final technique is motivated by our a posteriori observation that these techniques individually improve different sets of metrics, and therefore their combination may improve all metrics. In effect, the addition of random data acts to regularize the algorithm with large estimated redshift errors, so that it does not concentrate on only the worse performing target examples. 

{\rrr We note that we have also explored the use of a hybrid random selection method combined with targets which have  large estimated redshift offsets, but did not find a noticeable difference between this method and method 9)}.

We measure how well each targeting algorithm compares to the two base strategies by estimating the machine learning photometric redshift for a distinct sample of test galaxies (see \S\ref{obs_data} for details). The training data for this machine learning redshift estimation  is drawn from all galaxies up to and including the previous targeting run in each case.

\begin{table*}
\begin{center}
  \begin{tabular}{ | l | l |} 
 Acronym & Description \\ \hline
 SDSS & Select targets in the date order by which they were observed by the SDSS. \\ \hline
 Rand & Select targets at random. \\ \hline
LargeO & Select targets by the largest estimated redshift offset $b^{est}_{z_{ML}}$. \\ \hline
LargeE & Select targets by the largest estimated redshift error $\delta^{est}_{z_{ML}}$. \\ \hline
TsOE & Select targets by the largest estimated t-statistic: ($b^{est}_{z_{ML}}$/$\delta^{est}_{z_{ML}}$) \\ \hline
HmOE & Select targets with largest estimated harmonic mean:  ($b^{est}_{z_{ML}} \times \delta^{est}_{z_{ML}}$)/($b^{est}_{z_{ML}} + \delta^{est}_{z_{ML}}$) \\ \hline
TsOESamp & Select targets by drawing from the binned t-statistic distribution\\
& with a probability of being selected to be inversely proportional to the number of data in the bin. \\ \hline
HmOESamp & Select targets by sampling from the binned estimated harmonic mean distribution as for TsOESamp.\\ \hline
LE\_Rand & Select targets using a mixture of 50\% random (Rand) and 50\% of the largest estimated redshift error (LargeE).\\ \hline
\end{tabular}
\caption{\label{TargetSelectionCheatSheet} The different target selection routines explored in this work. The top two routines do not rely on machine learning, the other 7 routines use machine learning methods to estimate the redshift offsets $b^{est}_{z_{ML}}$, and the redshift errors $\delta^{est}_{z_{ML}}$, of the remaining targets. These values are then used in the specified way to select which targets should be subsequently observed in the next observing run.}
\end{center}
\end{table*}

\subsection{Simulated observing runs}
\label{setup}
We examine the effect of the target algorithms on the recovered redshifts of an independent test sample by constructing an experiment to answer these questions:

1) What is the optimal ordering that the target galaxies should have been observed in, if we had wanted to improve the redshift estimates on a {\rr representative} hold out sample.

2) How does this compare with the SDSS algorithm? We note that the SDSS algorithms were not designed with the above goal in mind. 

3) How do the target algorithms compare with a random selection algorithm?

We ensure that each experiment within SDSS I\&II or SDSS III have the same fixed pool of 100k target galaxies from which 1.3k galaxies are selected for each of the 75 simulated observing runs. Each algorithm will select targets from this fixed pool for follow up at different times. During each observing run we obtain a spectroscopic redshift for all of the selected target galaxies with 100\% success. {\rr This ignores important observation effects such as fiber collisions, seeing conditions, and airmass.} We commence each experiment after the first observing run has already been completed to provide an initial knowledge base to produce the initial machine learning predictions. We iteratively use all of the data acquired in previous runs to help decide which targets to observe in each subsequent run.  

We will end the experiment when all of the galaxies have been targeted in the simulated sets of observing runs, irrespective of the order in which they were targeted. Therefore at the end of the experiment all the targets will have been selected by each of the different targeting algorithms, and we expect all of the final machine learning redshift results to converge.

We have also explored other follow up survey configurations, such as increasing or decreasing the number of observing runs (including 50, 100, 200), or the number of spectra taken during each run. We find similar improvements in final results in these cases once similar amounts of spectra have been collected.

We finally compare the algorithms to determine which methods produce the quickest reduction in error when using the targeted galaxies as training galaxies in a standard machine learning redshift estimation process. This is performed by passing the independent test data through the trained machine systems to first predict redshifts, and then by measuring performance metrics.

\subsection{Estimating redshift performance on the test galaxies}
\label{metrics}
In the previous sections we determined which targets should be selected during the next hypothetical observing run. We measure how well these different sets of selected galaxies are able to predict machine learning redshifts of a final test sample. For this purpose we use two extremely large test samples; one for the SDSS I\&II analysis of size 769,479, and one for the SDSS III analysis of size 1,011,918, both of which are described in \S\ref{obs_data}. These large test samples enable very accurate estimates of the metrics as measured on the point prediction redshift residuals $\Delta_{z'} = (z_{spec}-z_{ML})/(1+z_{spec})$, of the following derived statistical quantities;  the median value $\mu$ of the distribution of $\Delta_{z'}$, the 68\% and 95\% spread of the distribution of $\Delta_{z'}$ defined as $\sigma_{68}(z'),\sigma_{95}(z')$ and the outlier fraction defined by the percentage of galaxies for which $|\Delta_{z'}|>0.15$. 

When presenting the results of the different targeting algorithms we measure and present the ratio of relative improvement in each of the metrics, with respect to the time ordered SDSS targeting selection criteria. A negative improvement implies that the targeted galaxies produce worse metrics than those obtained by the SDSS targeting order.

This choice of metric is unstable for values which are relatively small and oscillate around zero. This is the case for the median of the residual distribution, which is less than 0.004 at 95\%. We therefore do not examine the median value in what follows, and note that tree based methods often produce redshift estimates with low bias \citep[see e.g.][]{2008ASPC..394..521C,2015arXiv150308214H}.

%% file: results.tex
We begin by presenting the colour-magnitude distributions of galaxies which are selected by the different targeting algorithms with respect to the SDSS targeting algorithm. We then show how each targeting algorithm performs against the SDSS targeting order when estimating machine learning photometric redshifts of an independent test sample. We present results both as a function of the instantaneous value, corresponding to estimating redshifts as the survey is progressing, and as a function of the final value determined using all of the target galaxies. {\rr We then explore the effect of allowing a much enlarged target pool, which could mimic a change in follow-up strategy during the lifetime of the survey.}

\subsection{The distribution of targeted galaxies}
\label{disttargets}
\begin{figure*}
   \centering
   \includegraphics[scale=0.305,clip=true,trim=20 5 15 12]{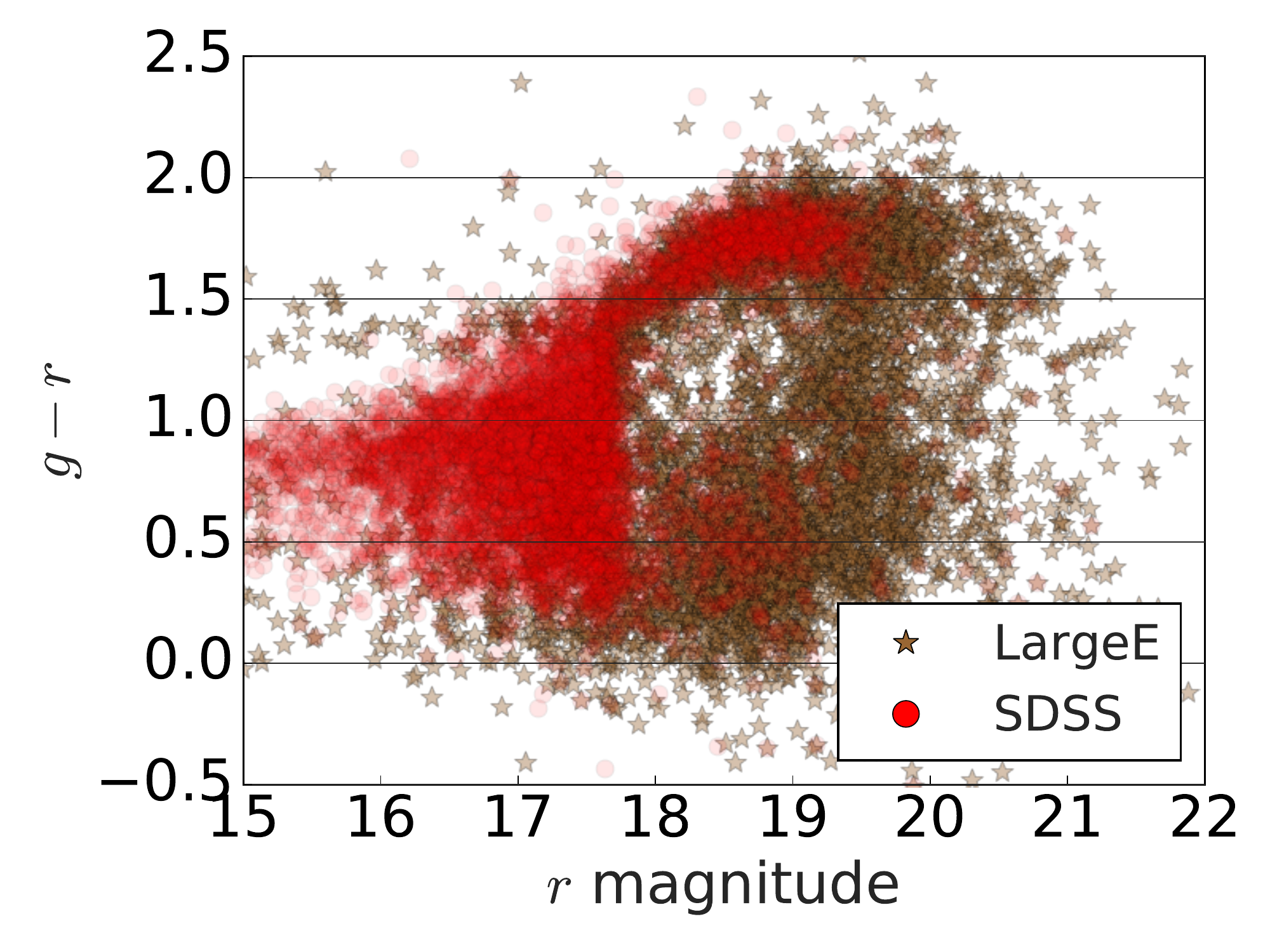}
   \includegraphics[scale=0.305,clip=true,trim=20 5 15 12]{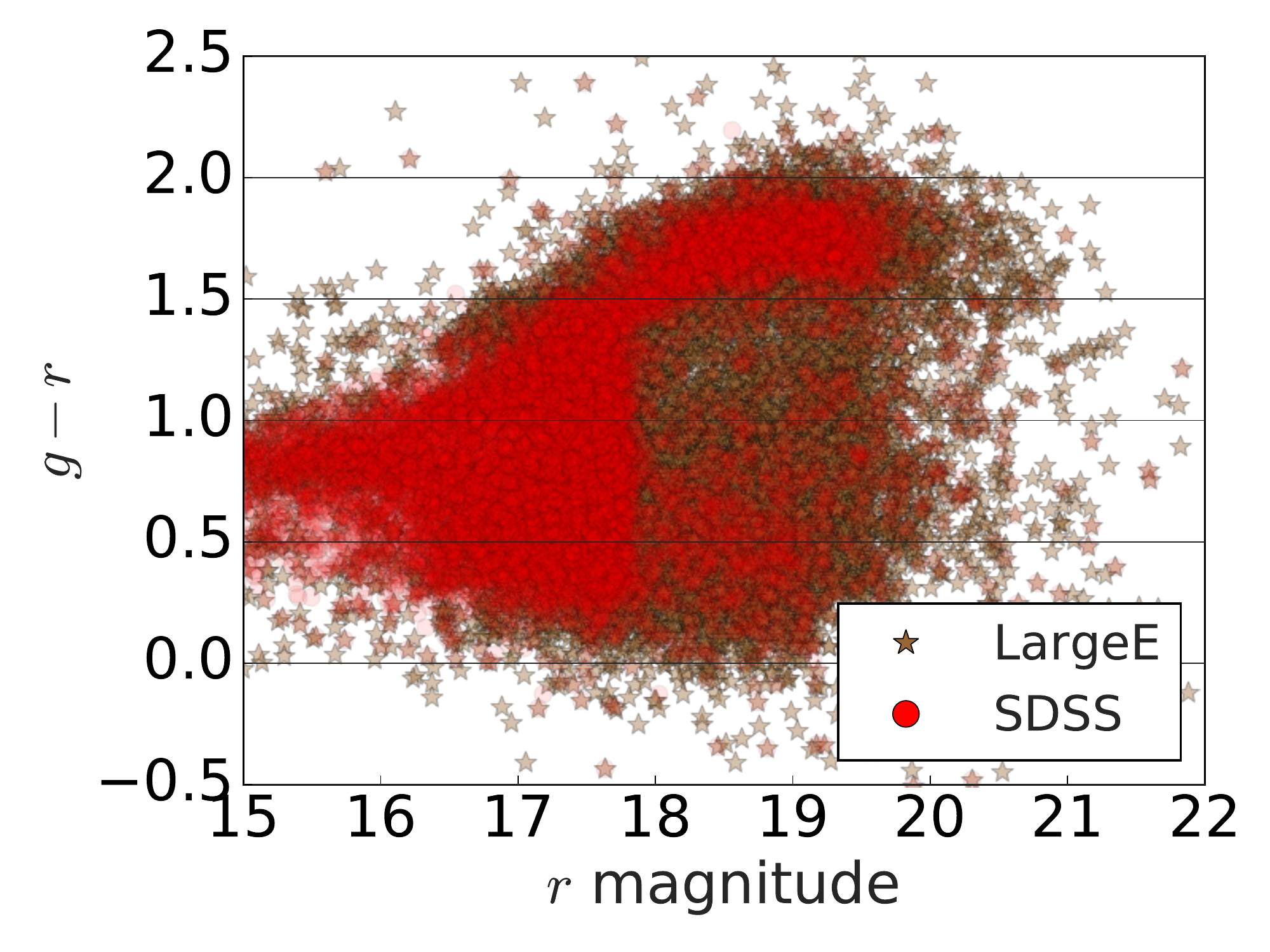}
   \includegraphics[scale=0.305,clip=true,trim=20 5 15 12]{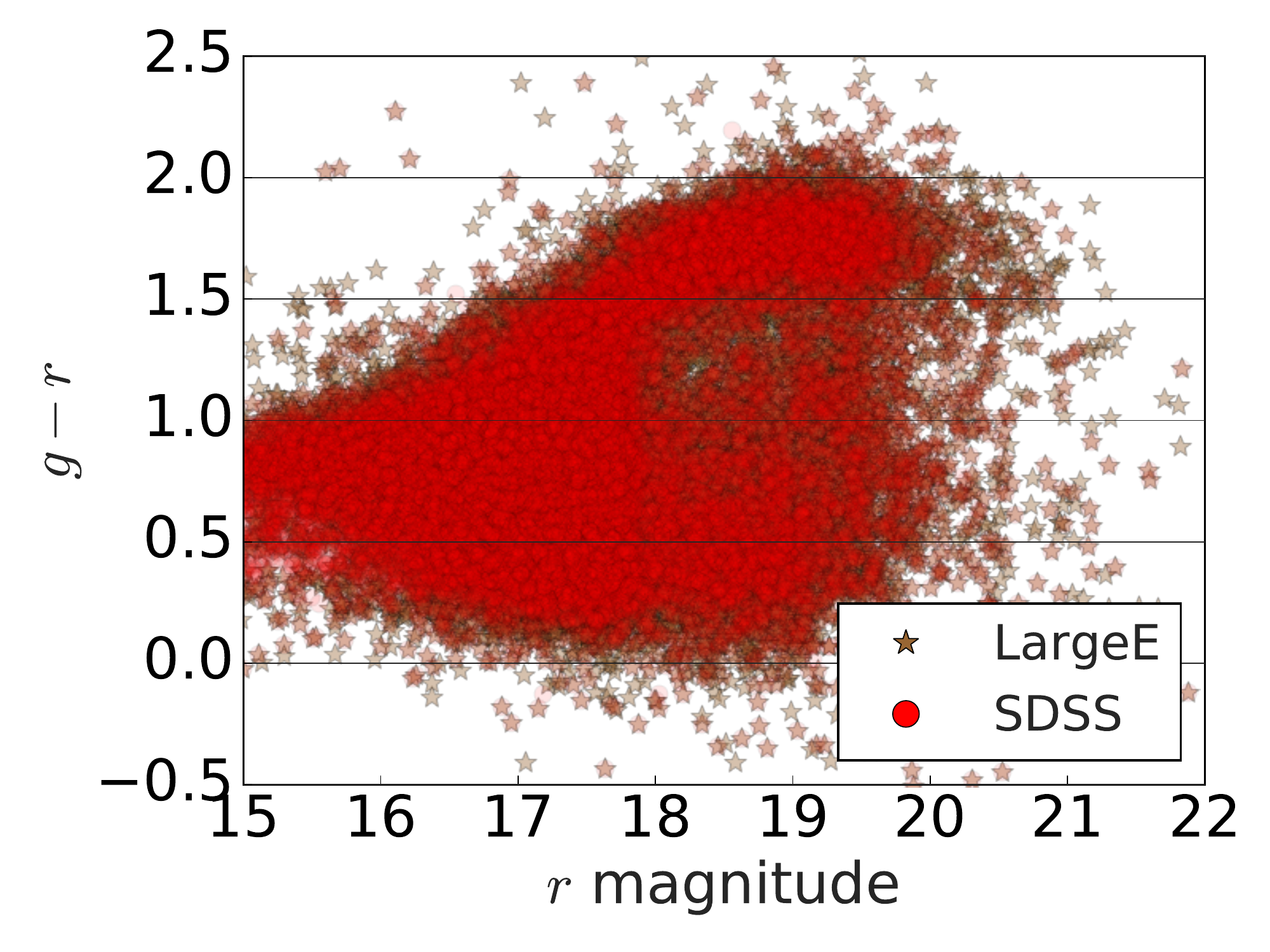}
   \caption{ \label{targettedGalaxies4} The {\rr $g-r$} colour magnitude distribution of SDSS I\&II galaxies selected by the SDSS and the Largest error (LargeE) target selection algorithms, after 5/75, 15/75, 35/75 (from left to right)  simulated  observing runs. Each observing run selects 1.3k galaxies.} 
\end{figure*}

In this section we examine which galaxies are preferentially selected by the different targeting algorithms. In Fig \ref{targettedGalaxies4} we show which galaxies are selected using the largest photometric error (LargeE) target selection algorithm, for an increasing number of simulated observing runs. The panels show the $g-r$ colour against $r$ band magnitude distribution after 5/75, 15/75 and 35/75 simulated sets of observations. We show the galaxies as observed in their original date order using the SDSS target algorithm by the red filled circles, and the LargeE targeting algorithm by the brown filled stars.  In this figure we use the SDSS I\&II data sample.

Examining the left-hand panel of Fig. \ref{targettedGalaxies4} we find that initially the two distributions of galaxies are very different, with the LargeE algorithm preferentially selecting fainter and bluer galaxies. The experiment has been constructed such that the final sample of targeted galaxies is the same. Therefore, as expected, we find that the different colour-magnitude distributions of galaxies become more similar as we progress through the survey. 

\begin{figure*}
   \centering
   \includegraphics[scale=0.305,clip=true,trim=20 5 15 12]{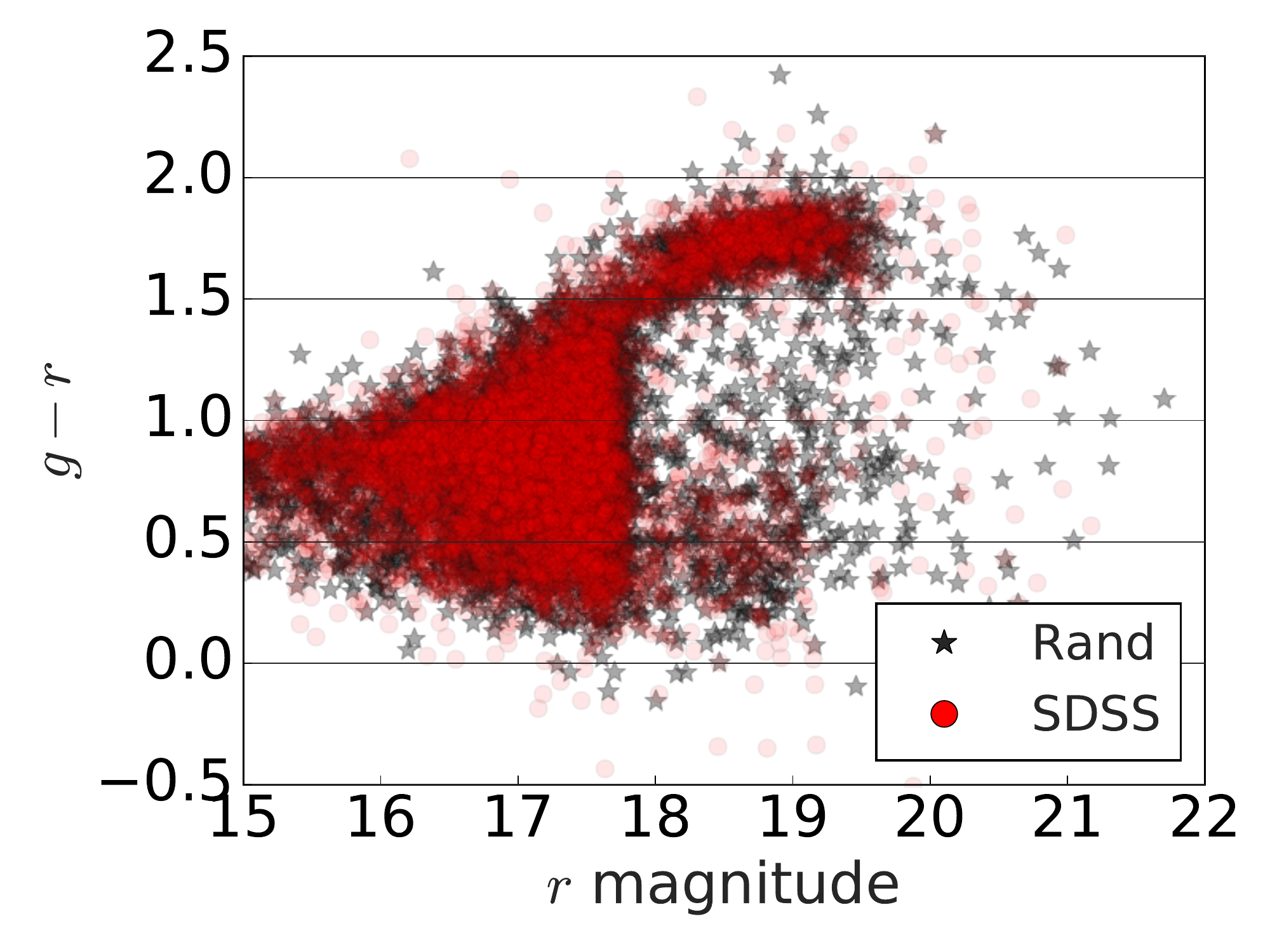}
   \includegraphics[scale=0.305,clip=true,trim=20 5 15 12]{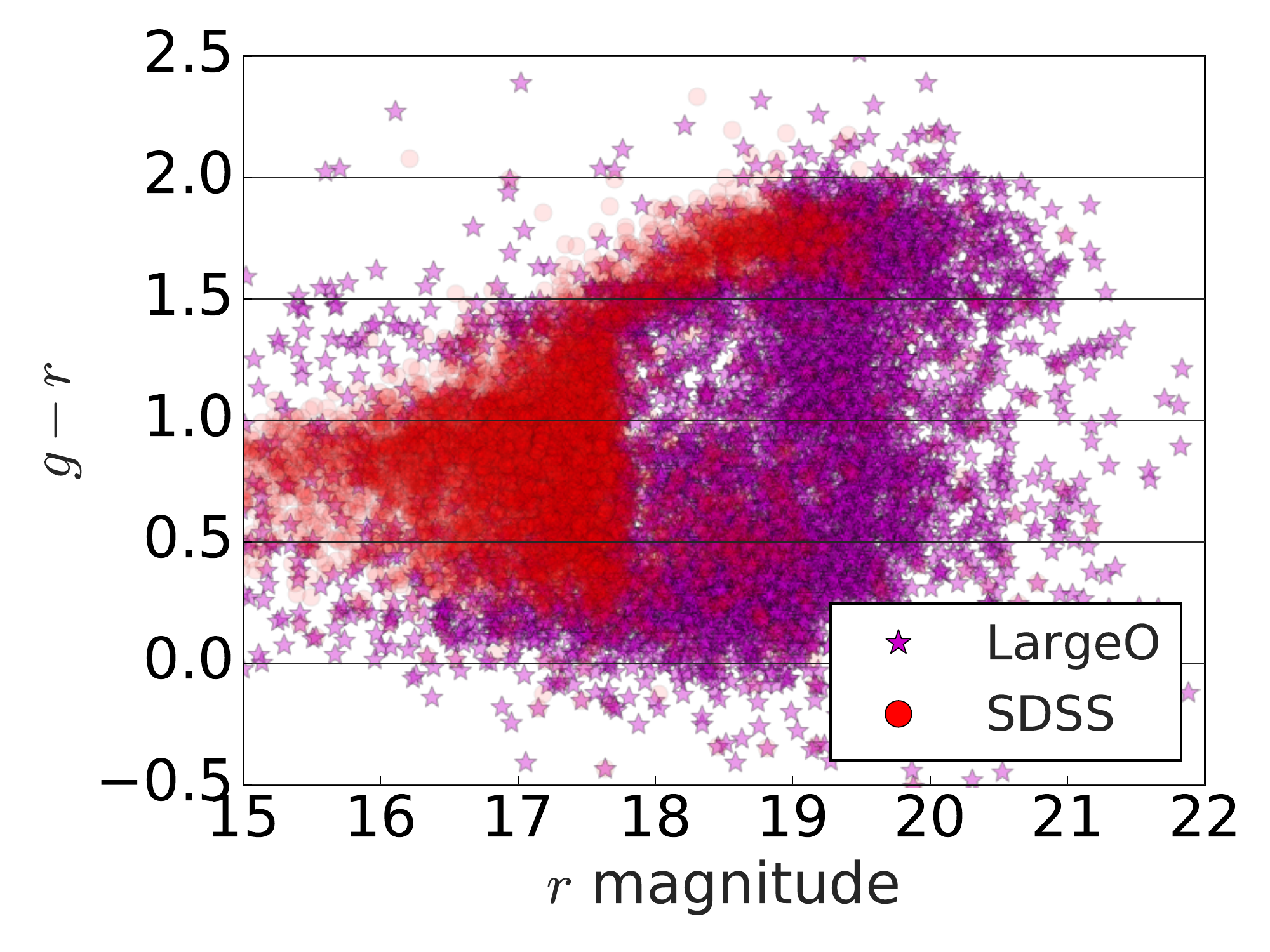}
   \includegraphics[scale=0.305,clip=true,trim=20 5 15 12]{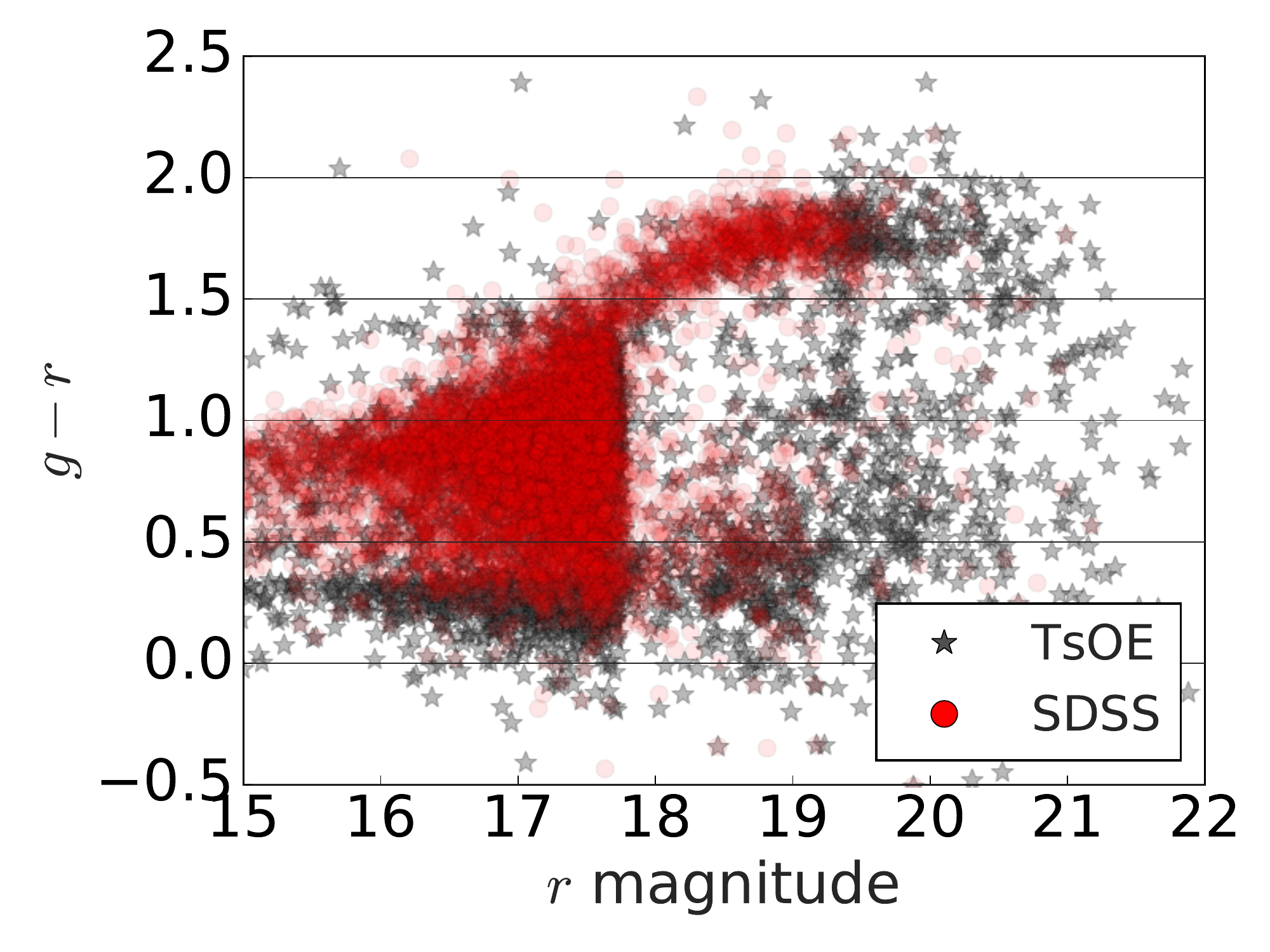}
   \caption{ \label{targettedGalaxies1} The {\rr $g-r$}  colour magnitude distribution of SDSS I\&II galaxies selected by the SDSS and the Random, Large predicted redshift bias, and large predicted t-statistic values. We show the different distribution of galaxies after 5/75 observing runs.} 
\end{figure*}

In Fig. \ref{targettedGalaxies1} we show snapshots after five observing runs of the same galaxy distribution as Fig. \ref{targettedGalaxies4} but for the following three different targeting algorithms from right to left: Random selection (Rand), Large Offset (LargeO), and the t-statistic (TsBE) between predicted bias and predicted error.

The panels of Fig. \ref{targettedGalaxies1} shows that the random targeting algorithm and the t-statistic distribution targeting algorithm (TsBO) populate different regions of colour-magnitude space. The algorithm TsBO preferentially selects $r$ band fainter galaxies across all colours, and a set of brighter, bluer galaxies. We note that the Large Bias algorithm closely resembles that of the Large Error targeting algorithm as seen in the left-hand panel of Fig. \ref{targettedGalaxies4}.

\begin{figure*}
   \centering
   \includegraphics[scale=0.305,clip=true,trim=20 5 15 12]{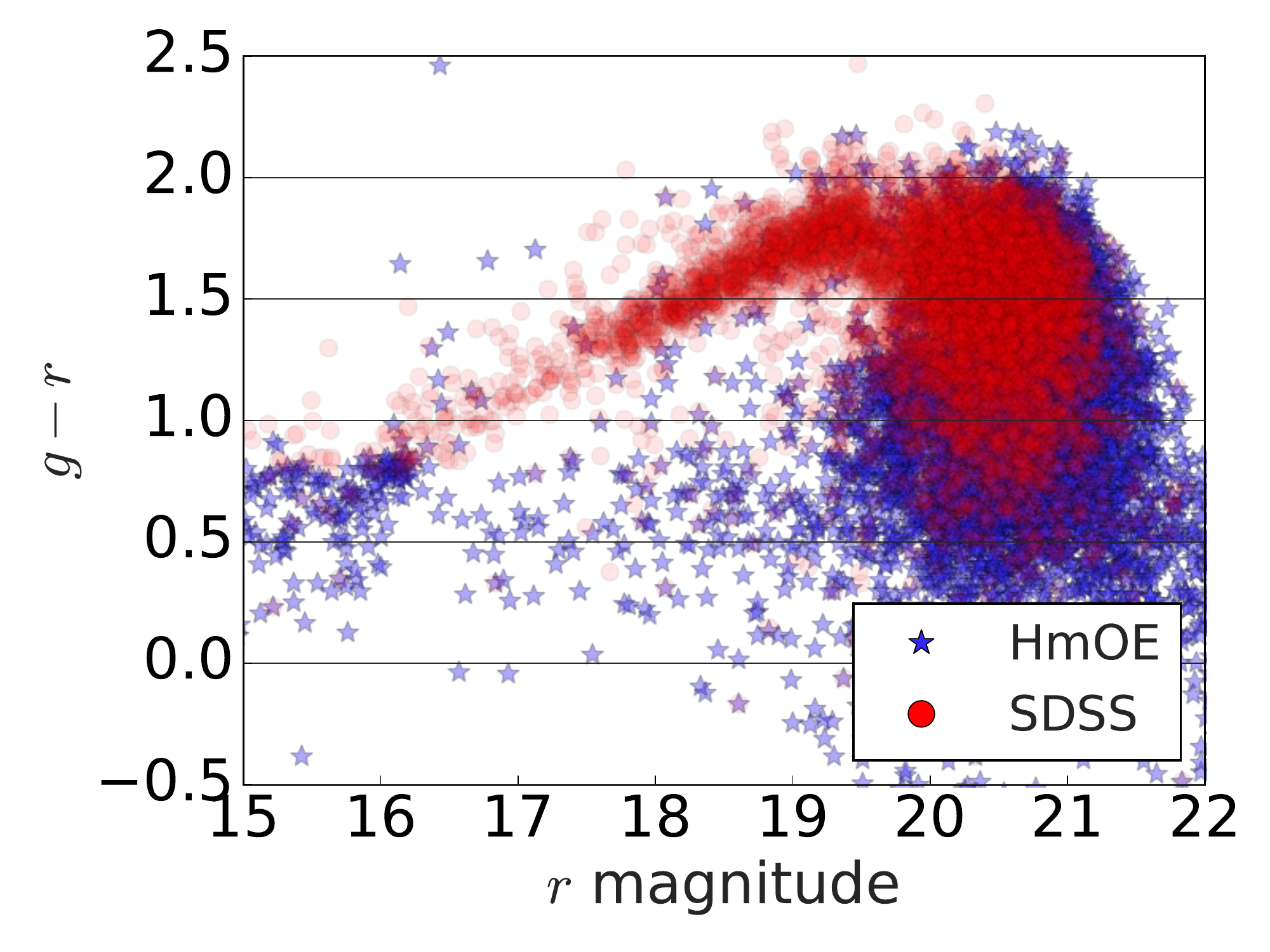}
   \includegraphics[scale=0.305,clip=true,trim=20 5 15 12]{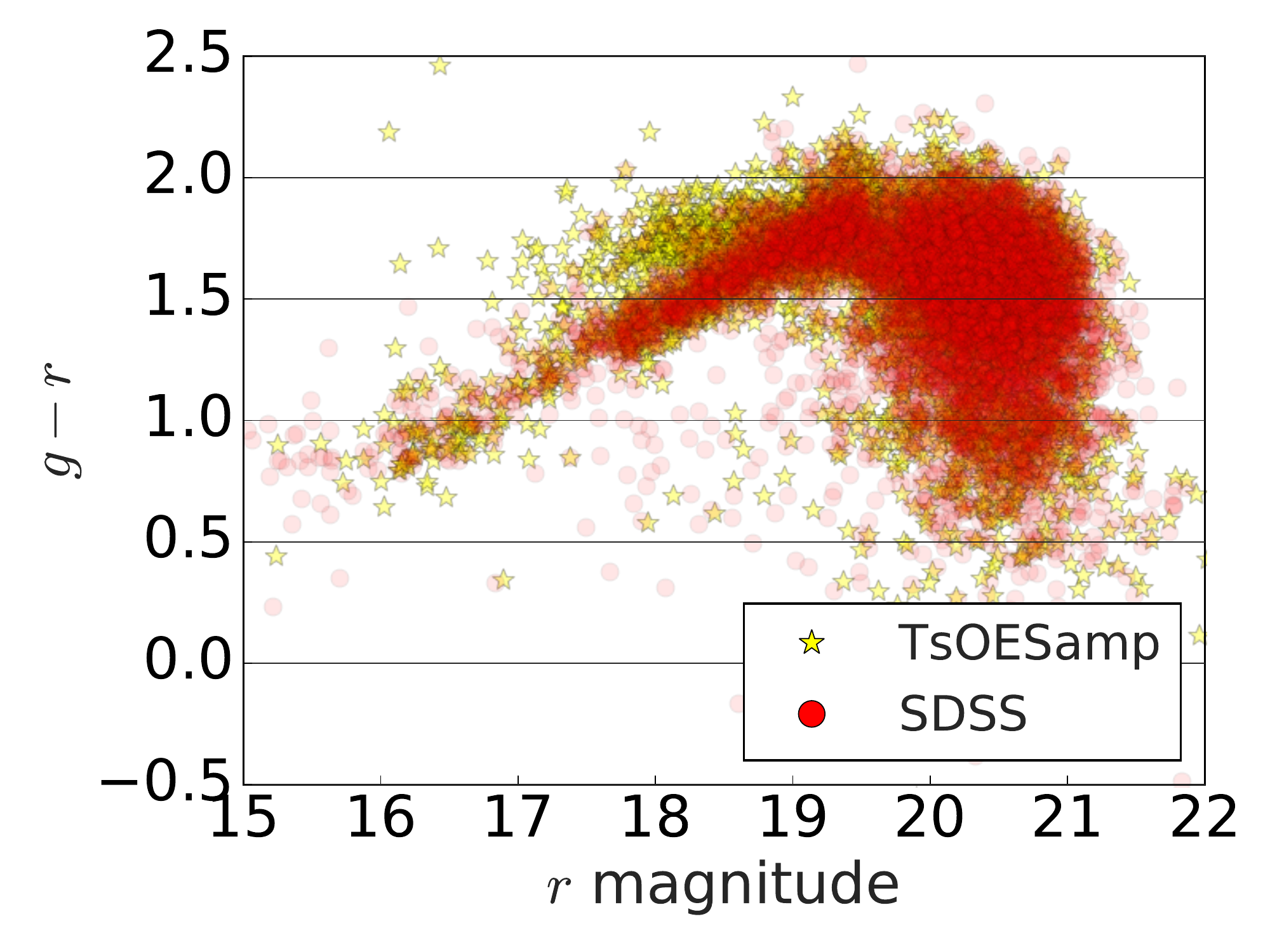}
   \includegraphics[scale=0.305,clip=true,trim=20 5 15 12]{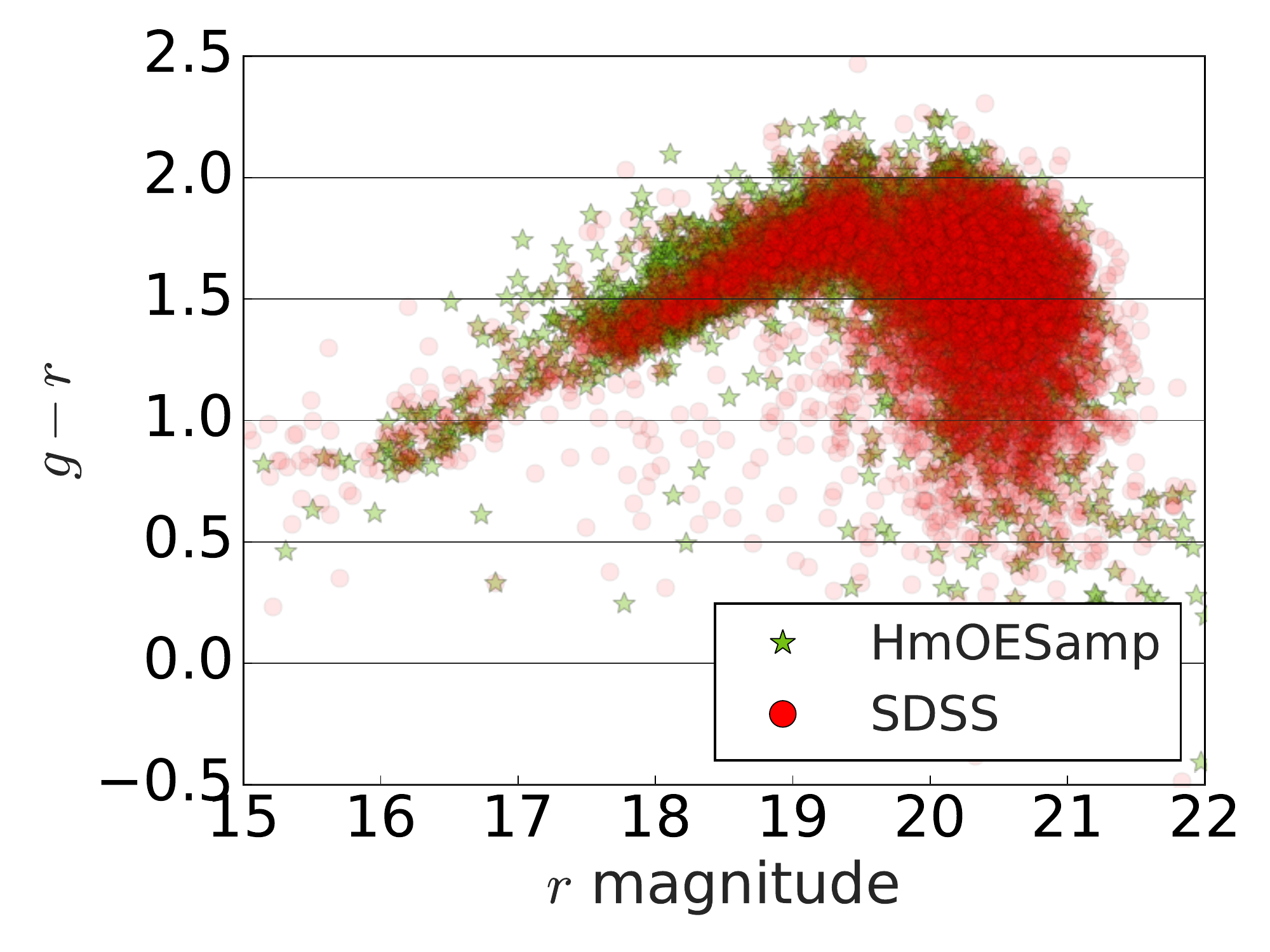}
   \caption{ \label{targettedGalaxies3} The {\rr $g-r$} colour magnitude distribution of SDSS III galaxies selected by the SDSS and the HmBE, TsBESamp, and HmBESamp algorithms (from left to right) after 5/75  simulated observing runs.} 
\end{figure*}

In Fig. \ref{targettedGalaxies3} we choose to show the colour-magnitude distribution for the SDSS III sample, instead of the SDSS I\&II sample as in the previous figures, of galaxies for the final three different targeting algorithms: the Harmonic mean of the predicted bias and predicted error (HmBE), the inverse number sampled t-statistic distribution TsBESamp, and the inverse number sampled Harmonic mean distribution (HeBESamp).

Examining Fig. \ref{targettedGalaxies3} we find that both the sampled t-statistic and sampled Harmonic mean targeting algorithms produce samples of galaxies which are very similar to the inherent SDSS ordering. The Harmonic mean {\rr HmOE} targeting algorithm preferentially selects bluer galaxies at all $r$ band magnitudes. 

{\rr We note that in all of the figures in this section there are clearly different color-magnitude populations of galaxies being selected. These correspond to the different observing goals of both SDSS I\&II and SDSS III, and correspond to main sample galaxies, luminous red galaxies, and quasars. 
}

\begin{figure*}
   \centering
   \includegraphics[scale=0.46,clip=true,trim=20 5 15 12]{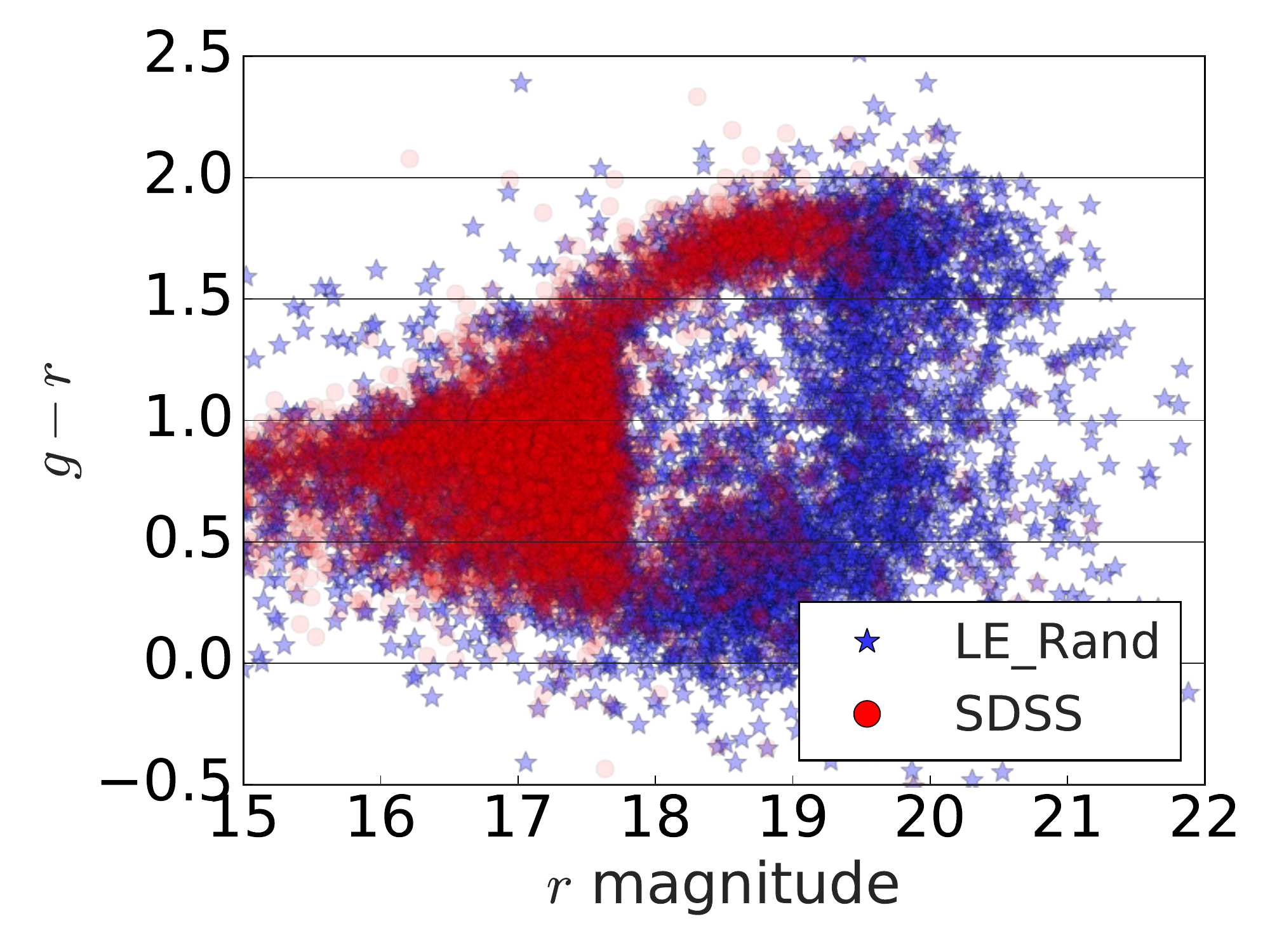}
   \includegraphics[scale=0.46,clip=true,trim=20 5 15 12]{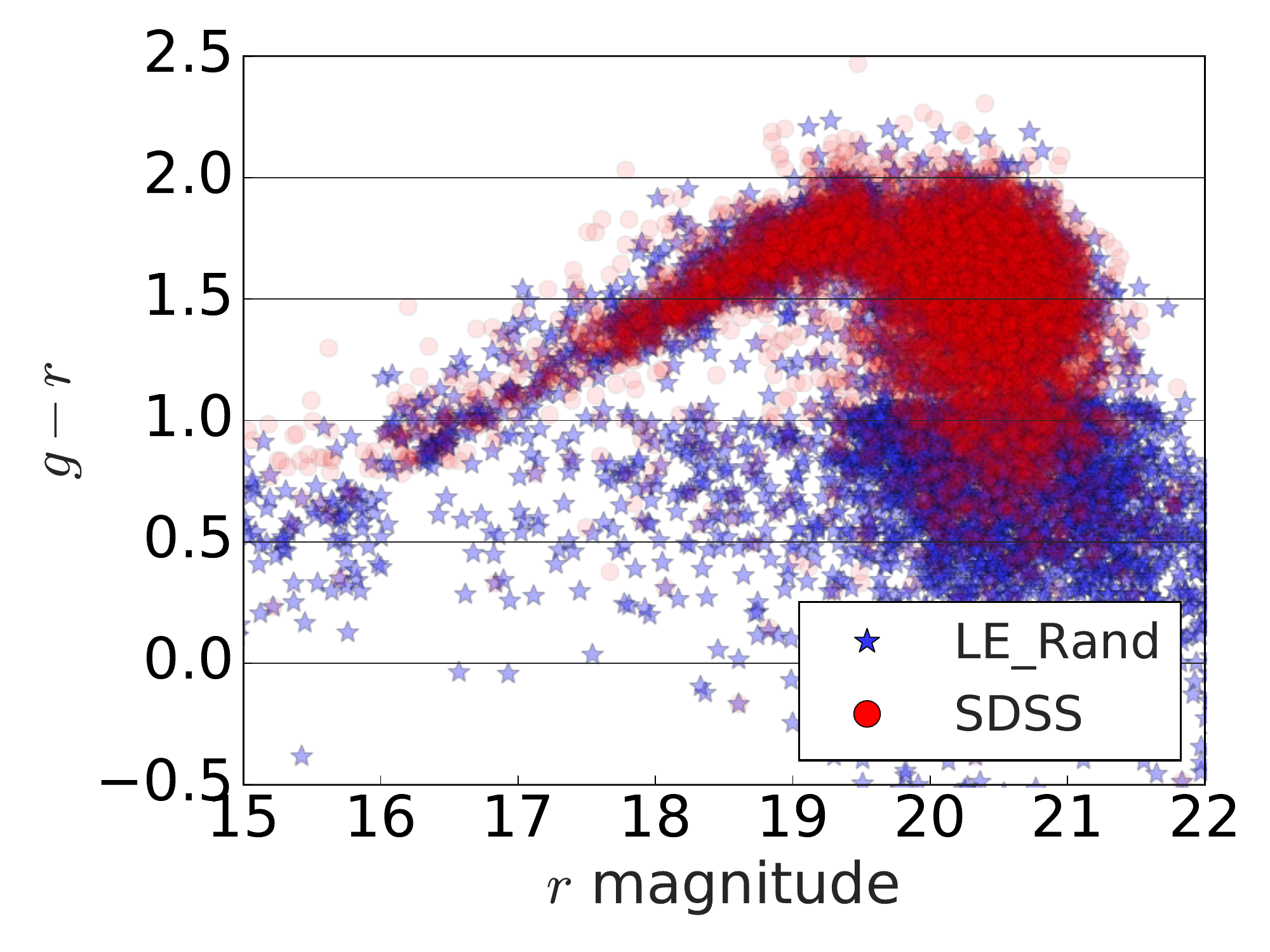}
   \caption{ \label{targettedGalaxies5}
   The {\rr $g-r$} colour-magnitude distribution of SDSS I\&II (left-hand panel) and SDSS III (right-hand panel) galaxies using the SDSS and the hybrid LE\_Rand target selection algorithms. We again show the distribution of galaxies after 5/75 simulated observing runs.}
\end{figure*}

Finally in Fig. \ref{targettedGalaxies5} we show the colour-magnitude distribution of galaxies using the hybrid random and large predicted redshift error (LargeE\_Rand) algorithm for both the SDSS I\&II (left-hand panel) and SDSS III (right-hand panel) after five simulated  observing runs.
In Fig. \ref{targettedGalaxies5} we see that the colour-magnitude distribution of galaxies resembles that of both the random target selection (see the left-hand panel of Fig. \ref{targettedGalaxies1}), and the targeting algorithm using the predicted redshift error (see the left-hand panel of Fig. \ref{targettedGalaxies4}).

\subsection{The performance of different targeting algorithms}
\label{disttargets}
\begin{figure*}
   \centering
   \includegraphics[scale=0.46,clip=true,trim=0 15 40 20]{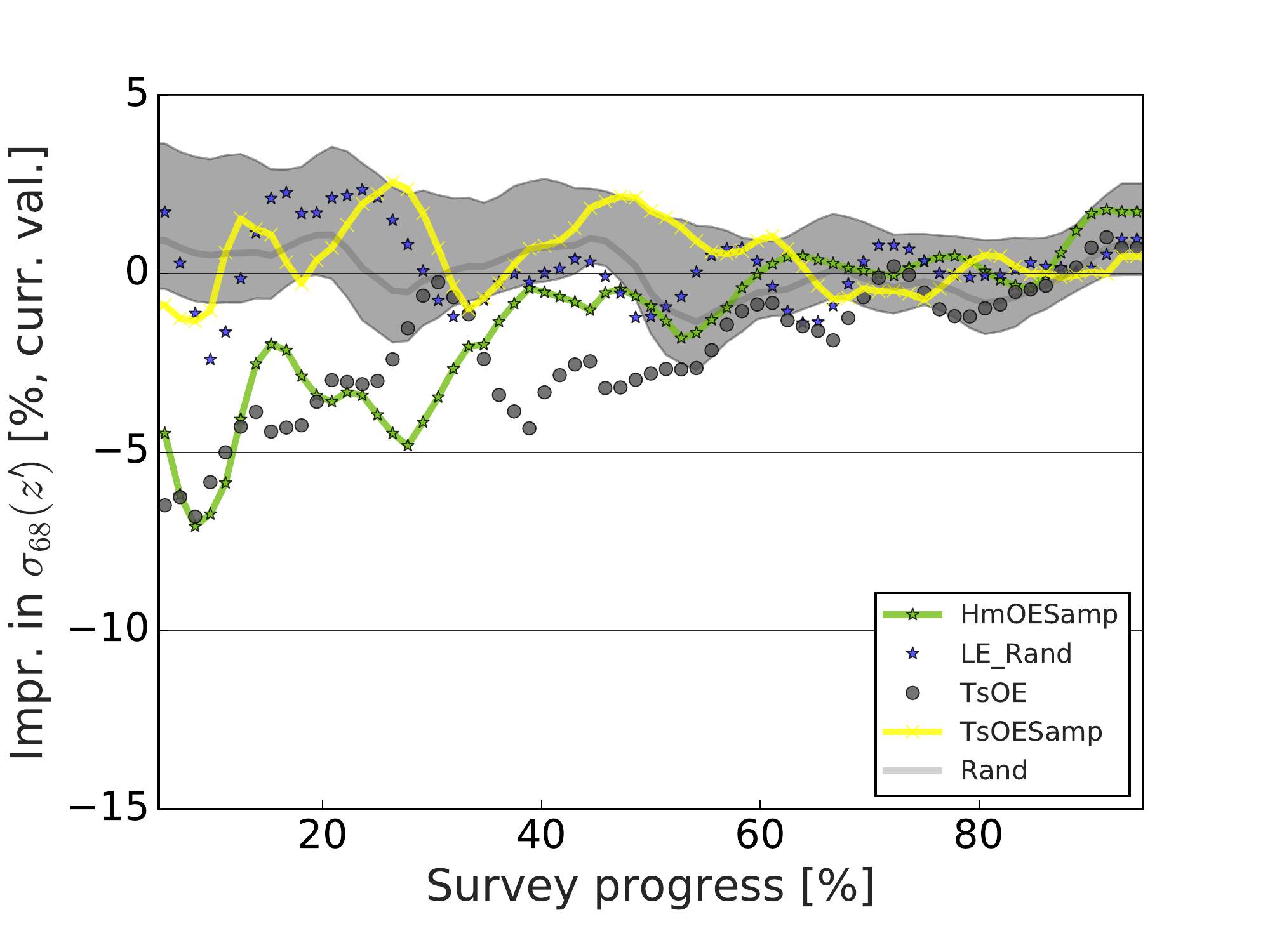}
   \includegraphics[scale=0.46,clip=true,trim=0 15 40 20]{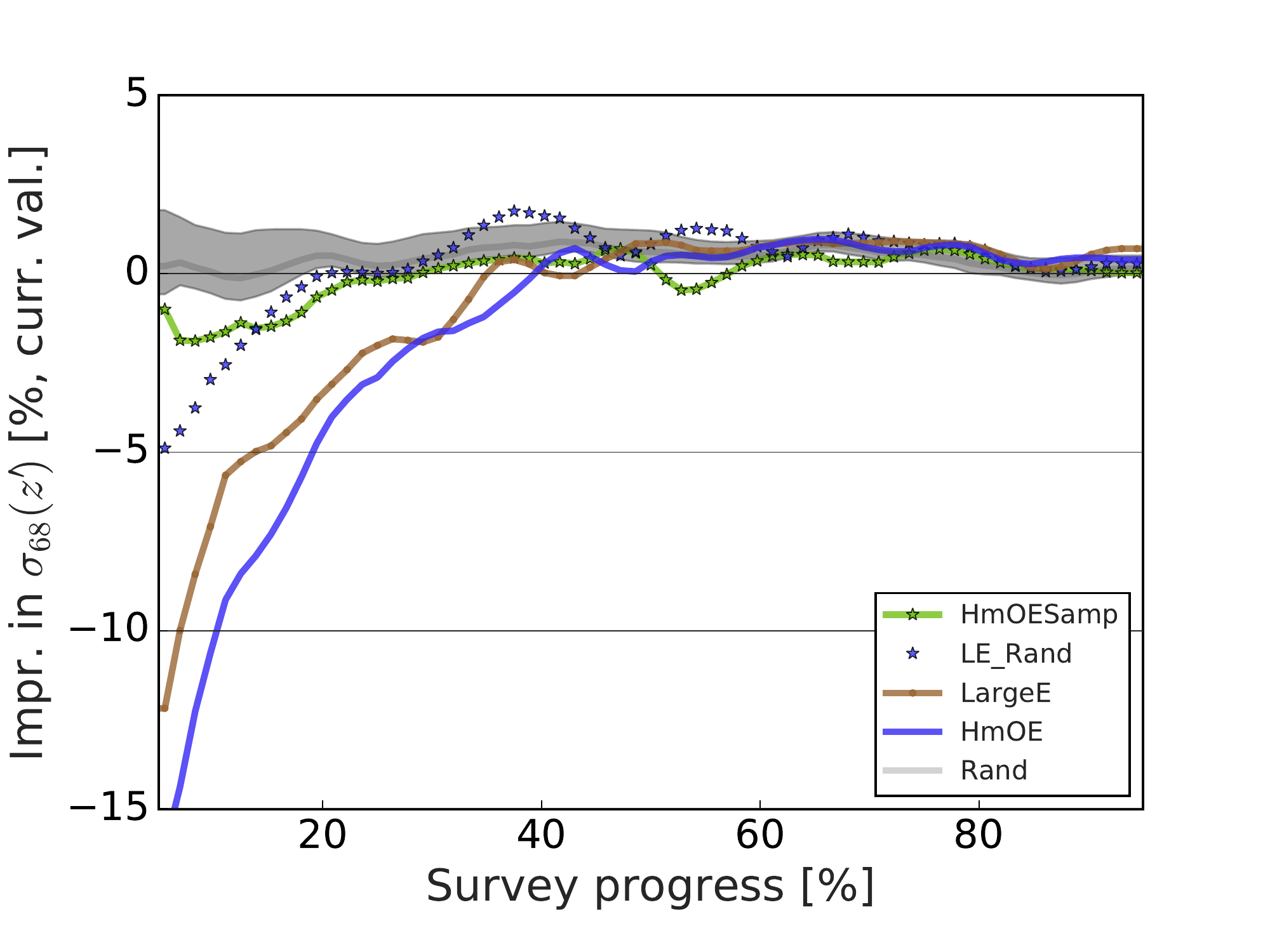}\\
   \includegraphics[scale=0.46,clip=true,trim=0 15 40 20]{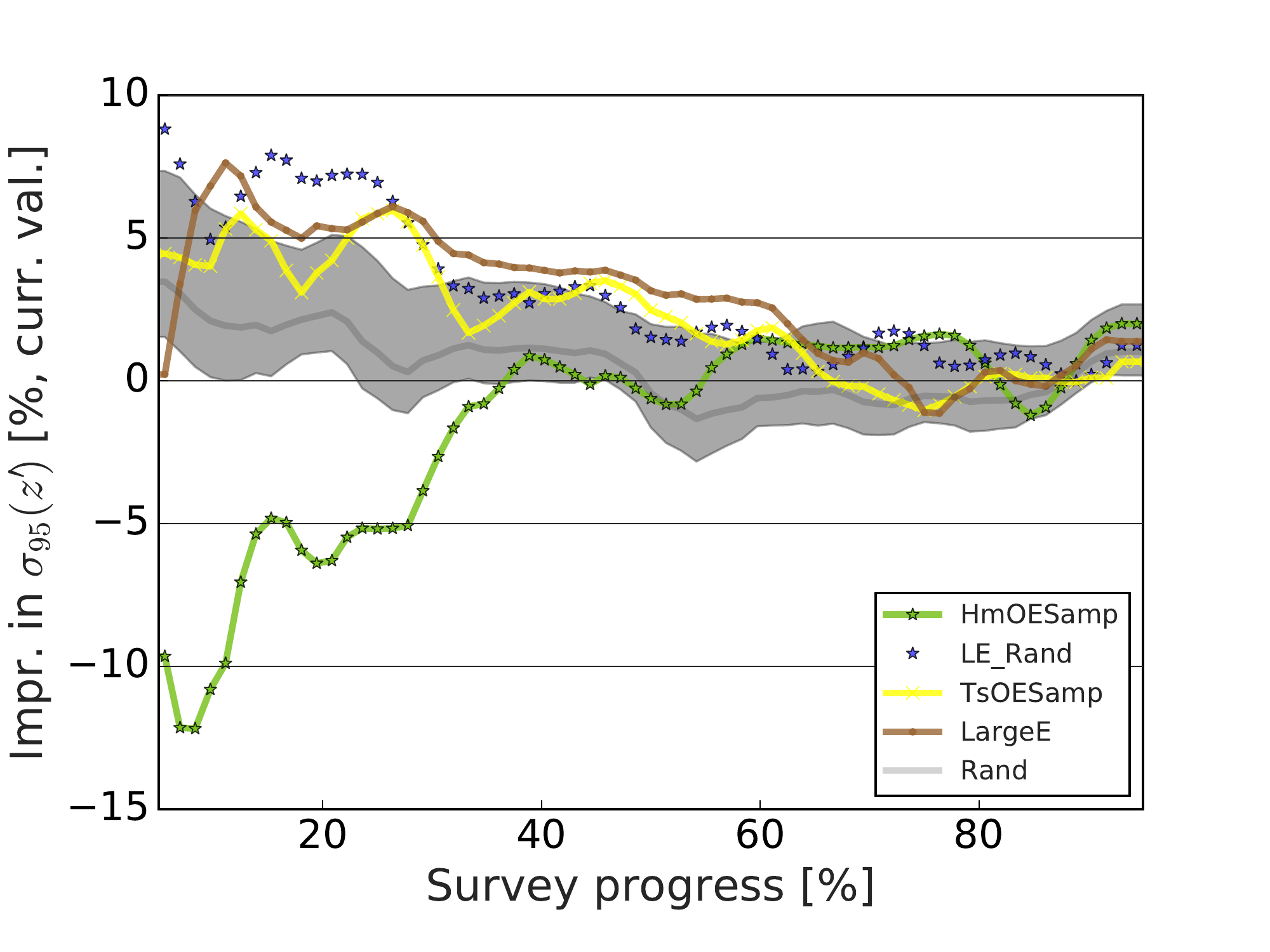}
   \includegraphics[scale=0.46,clip=true,trim=0 15 40 20]{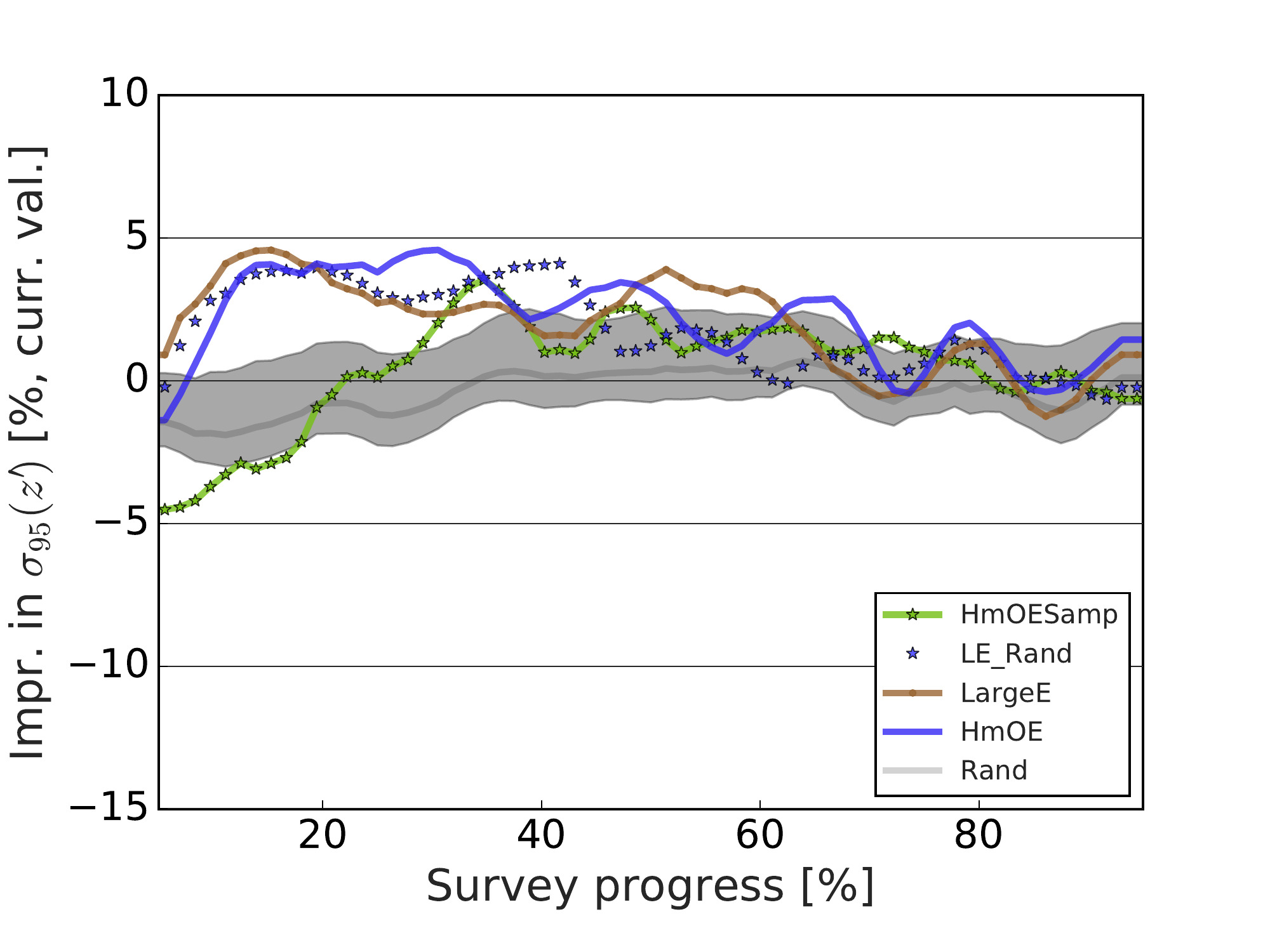}\\
   \includegraphics[scale=0.46,clip=true,trim=0 15 40 15]{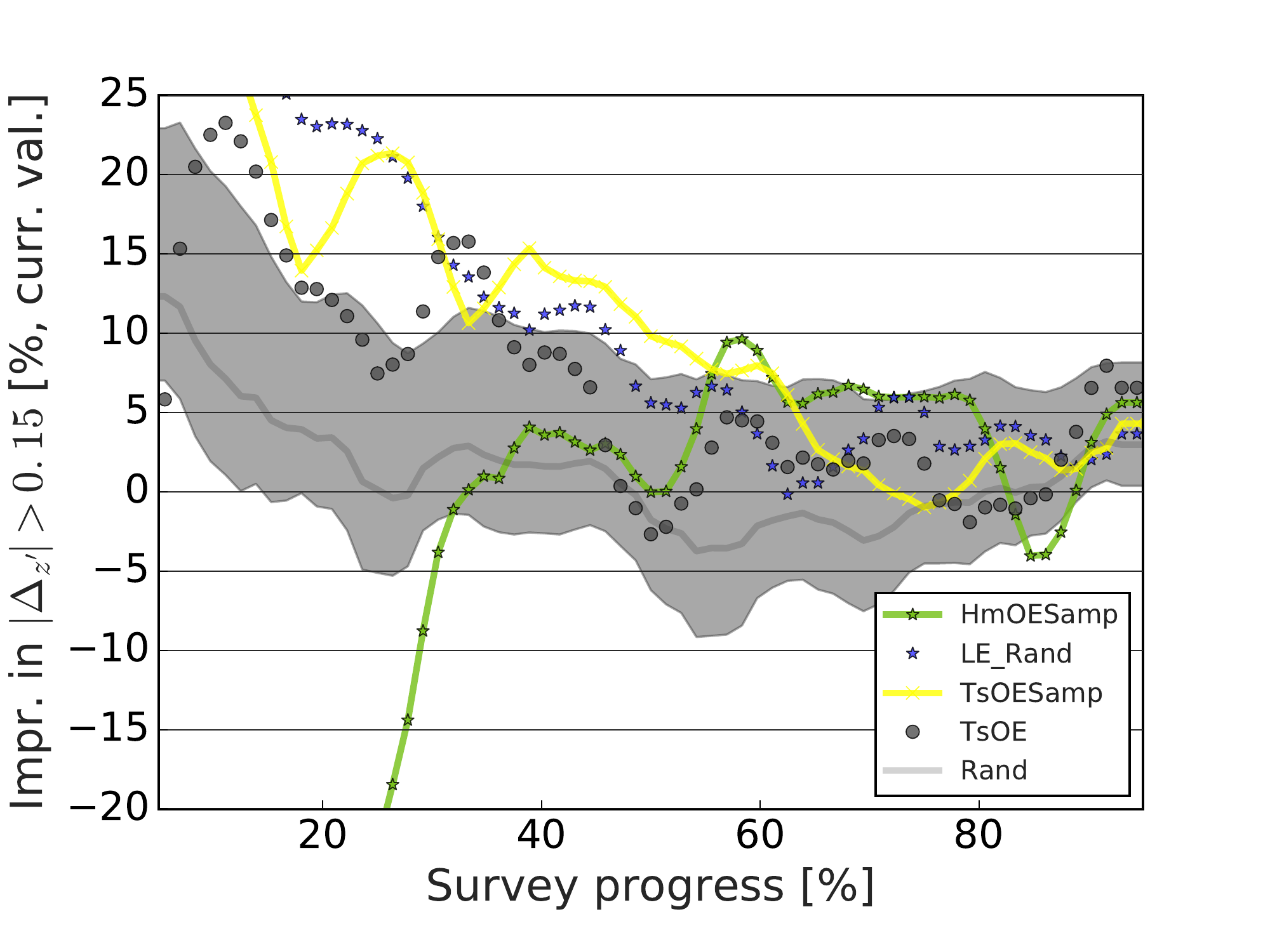}
   \includegraphics[scale=0.46,clip=true,trim=0 15 40 15]{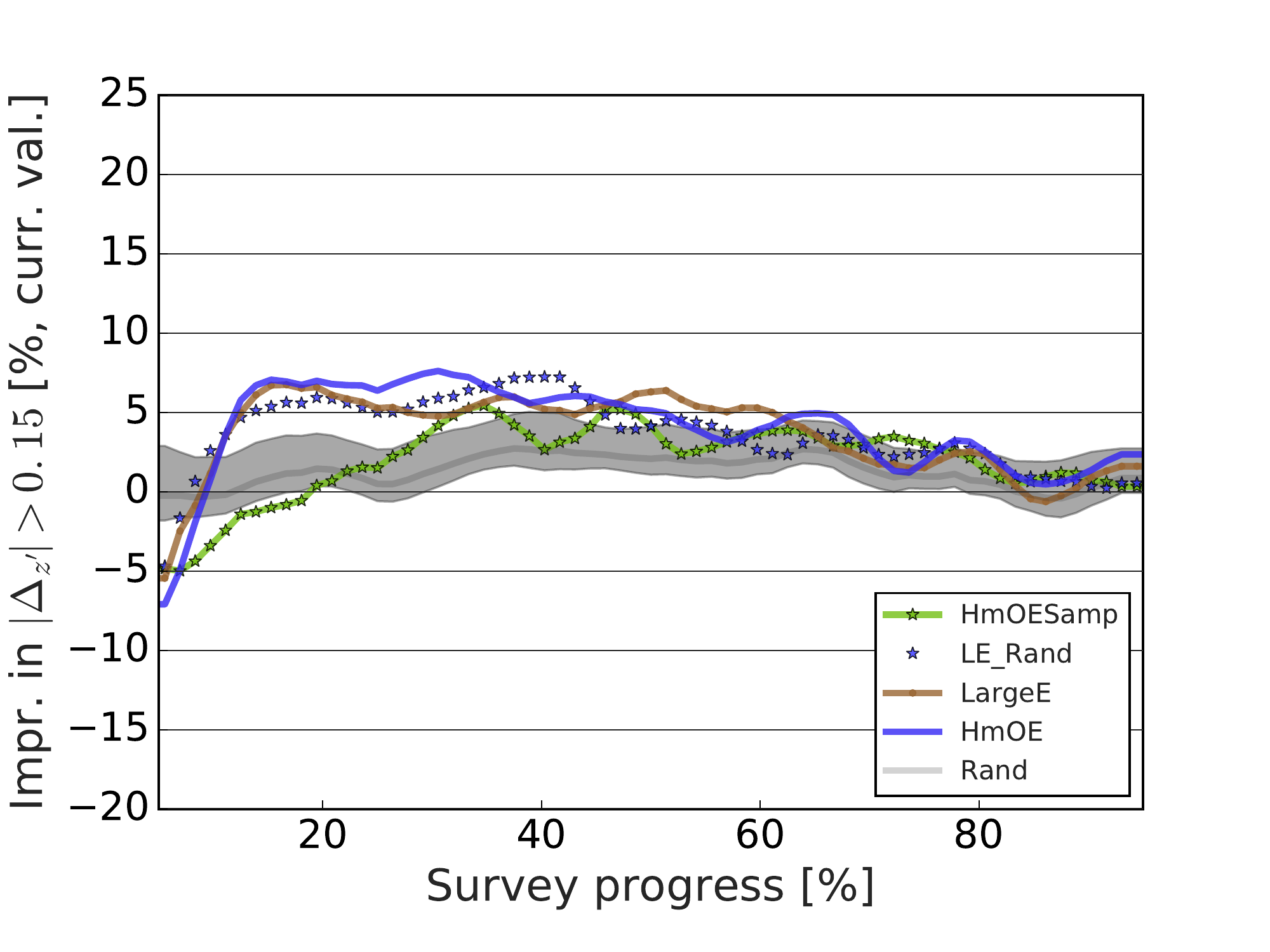}
   \caption{ \label{targetSelectionEffect1} The relative improvement of each measured metric (see y-axis label) value for different targeting algorithms with respect to the instantaneous (or current, curr.) value from the SDSS time ordered targeting algorithm. The left-hand (right-hand) panels show the results for the SDSS I\&II (SDSS III) analyses. The shaded region corresponds to the 68\% spread of metric values using the random targeting algorithm (Rand) across 7 independent experiments.} 
\end{figure*}

We show the effect of the different targeting algorithms using the metrics $\sigma_{68}(z'),\sigma_{95}(z'),|\Delta_{z'}|>0.15$ to describe photometric redshift performance as calculated on an independent test sample in Fig. \ref{targetSelectionEffect1} and Fig \ref{targetSelectionEffect2}. We show SDSS I\&II analysis in the left-hand panels and the SDSS III analysis in the right-hand panels. We show the relative improvement in each of the metrics, with respect to the time ordered SDSS targeting selection algorithm. We measure the relative improvement in each of these metrics with respect to two different SDSS values. In Fig. \ref{targetSelectionEffect1} we show the relative improvement with respect to the instantaneous SDSS metric value, as calculated using the sample of galaxies which have been targeted up to and including that number of targeting runs. In  Fig \ref{targetSelectionEffect2} we show the relative improvement with respect to the final value of the SDSS metric, which is calculated using the full sample of targeted galaxies, and is approximately the same as the final value from all other algorithms. In both figures the shaded contours around the random targeting algorithm (Rand) shows the 68\% spread of that metric over 7 different realisations, and gives an estimate of the expected error on the other targeting routines. In each figure we always show the benchmark (SDSS), and (Rand) algorithms, and the overall best (LE\_Rand) and worst (HmBESamp) performing algorithms, and then a random selection of two other algorithms. This choice is made for aesthetic considerations. We discuss each figure below.

\subsubsection{Comparison with the instantaneous SDSS metric values}
\label{inst}
In the top panels of Fig. \ref{targetSelectionEffect1} we see that all of the targeting algorithms except for TsBESamp and LE\_Rand perform less well than the random selection algorithm on the statistic $\sigma_{68}(z')$ as measured on the independent redshift scaled residuals of the final test sample. This suggests how these results may apply to representative and unseen data. {\rr We note that the targeting algorithms not shown behave similarly to those presented.}

The middle and bottom panels of Fig. \ref{targetSelectionEffect1} show how the targeting algorithms perform on the metrics $\sigma_{95}(z')$  and $|\Delta_{z'}|>0.15$ as calculated on the independent test sets. We find that almost all of the target selection algorithms either slightly, or moderately, out perform the random selection algorithm after approximately 20\% of the simulated observing runs have been completed. This improvement can be expected because many of the targeting algorithms, including LargeE, select galaxies which are predicted to have large estimated values of redshift offset and redshift error, see also \S\ref{selection}. By preferentially selecting these galaxies we expect the wings of the distributions of $\Delta_{z'}$ to be removed. This also explains why the improvement with respect to the random algorithm increases as we pass from $\sigma_{95}(z')$  to $|\Delta_{z'}|>0.15$, which effectively samples a larger component of the wings of the distribution. 

Finally we note that as the survey surpasses the 60\% completion mark, all of the algorithms converge. We can understand this by recalling that this experiment has been performed by imposing a fixed list of potential targets, and each algorithm will eventually have selected the same target galaxies. We explore this effect further in \S \ref{moredata} by relaxing this constraint. We note that in all panels in Fig. \ref{targetSelectionEffect1} we find that the targeting algorithm HmBESamp performs worse than all the other sampling algorithms including that of the random sampling (Rand). 

Examining all panels of Fig. \ref{targetSelectionEffect1} we see that the random targeting algorithm is always close to 0\% improvement compared to the SDSS, which implies consistency with the measured value of each metric between both algorithms. This suggests that the SDSS target algorithm selected galaxies randomly once target lists were made available to them. This conclusion is consistent with their targeting strategy \citep[see][]{2002AJ....124.1810S}.

If we concentrate on the LE\_Rand algorithm in each panel of Fig. \ref{targetSelectionEffect1}, we see that for a given number of target runs, this  targeting algorithm either does no worse than, or improves by 5-25\%, the measured metrics when compared with the SDSS and Rand targeting algorithms. This motivates the a priori choice for the construction of this targeting algorithm made in \S\ref{selection}. This enhancement is less pronounced for the SDSS III sample, but there is still improvement with respect to both the SDSS and the random targeting selection algorithms. The distribution of SDSS III galaxies is inherently redder and more homogeneous than that of the SDSS I\&II, e.g., see Fig. \ref{targettedGalaxies5}. This explains why the improvement in the metrics is less pronounced as more data is added to the machine learning systems. We conclude that if the SDSS had wanted to optimise their target selection algorithm,  to improve a machine learning redshift estimation of a final test sample, {\rr {\it and} the full list of potential targets were available apriori,} they would have been able to improve their redshift metrics by 5-25\% using the suggested LE\_Rand selection algorithm. These improved redshift estimates could have potentiality lead to earlier science results for some science applications.

\subsubsection{Comparison with the final SDSS metric values}
\label{final}
\begin{figure*}
   \centering
   \includegraphics[scale=0.46,clip=true,trim=0 15 40 30]{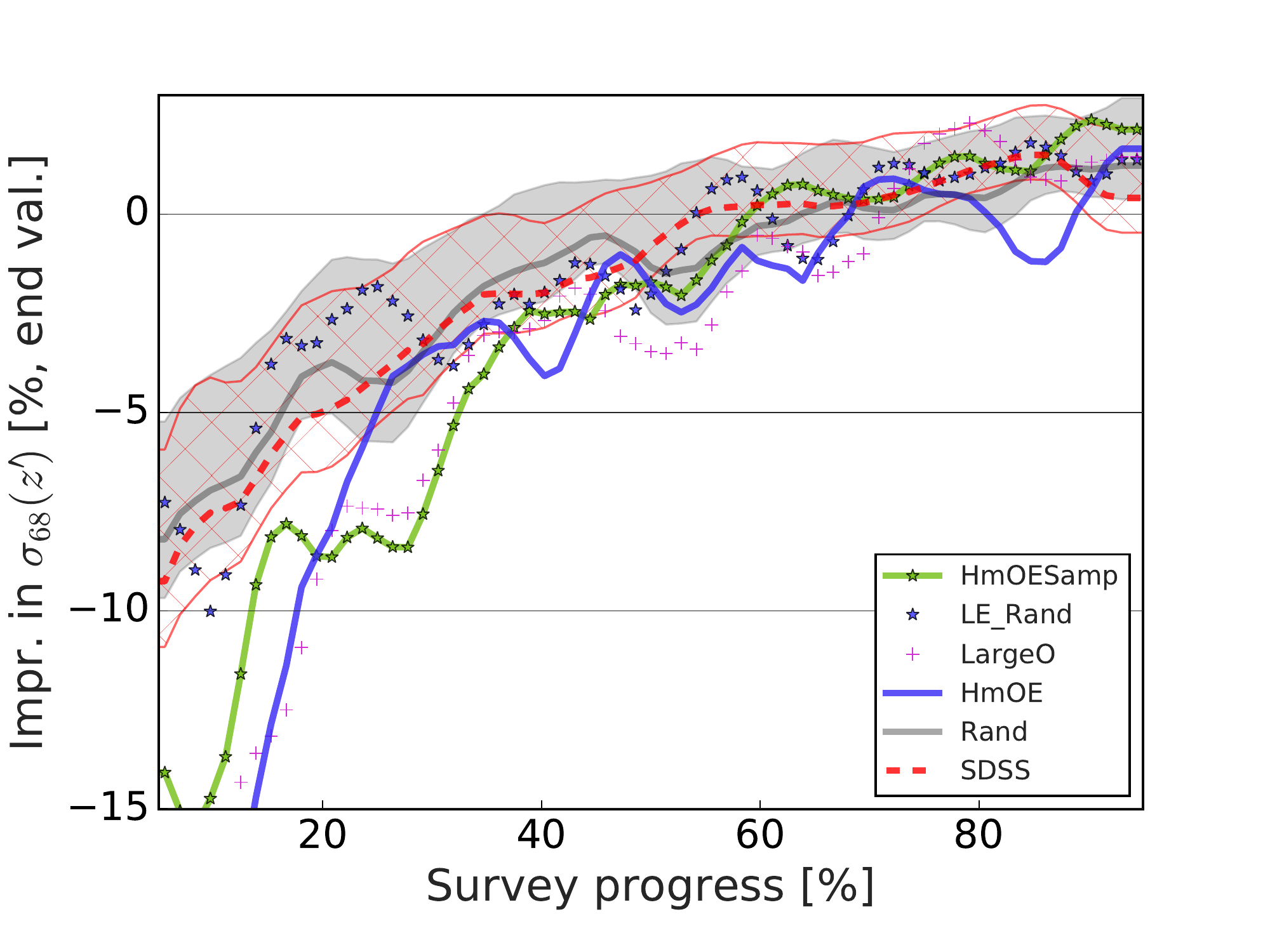}
   \includegraphics[scale=0.46,clip=true,trim=0 15 40 30]{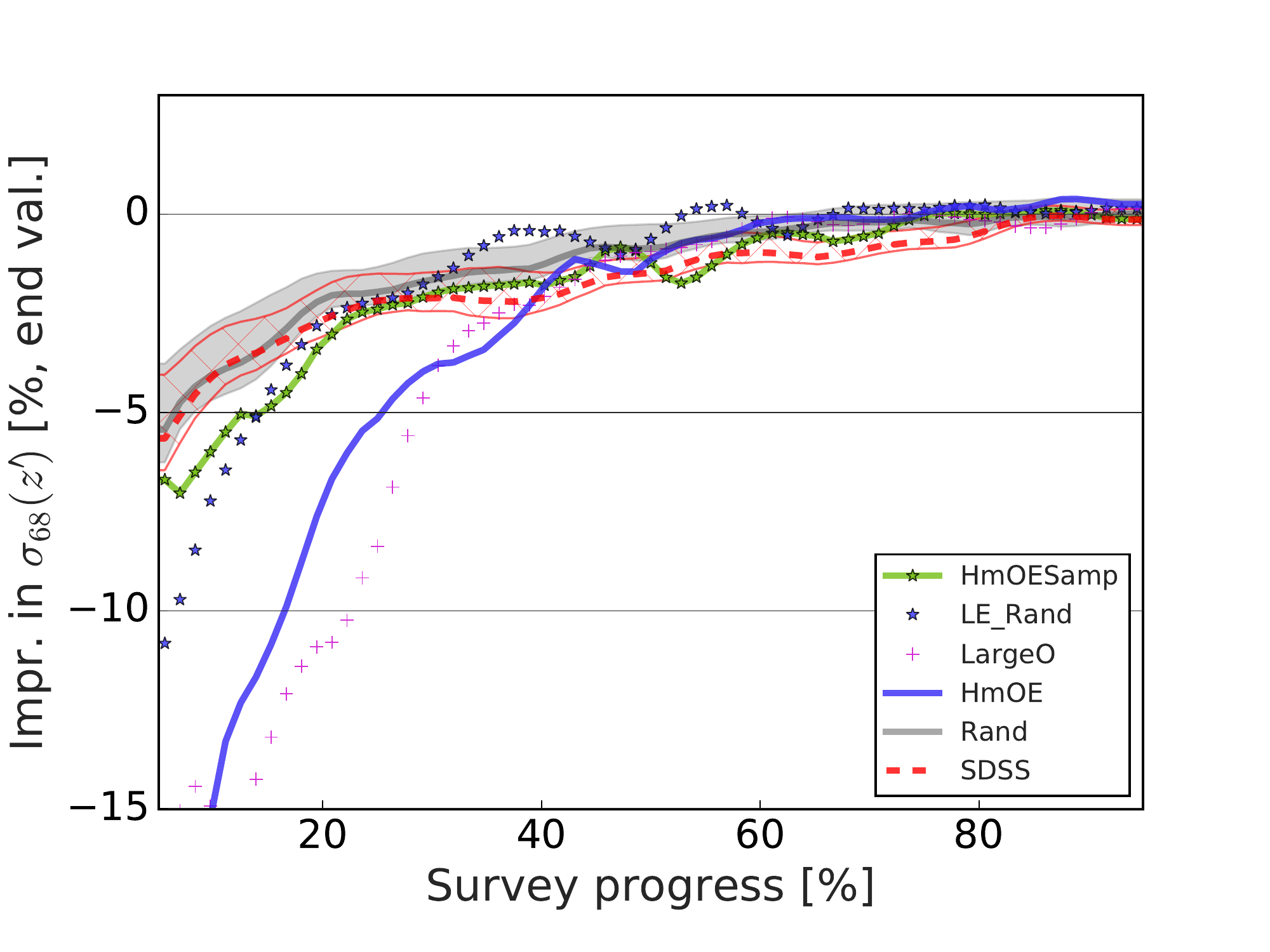}\\
   \includegraphics[scale=0.46,clip=true,trim=0 15 40 30]{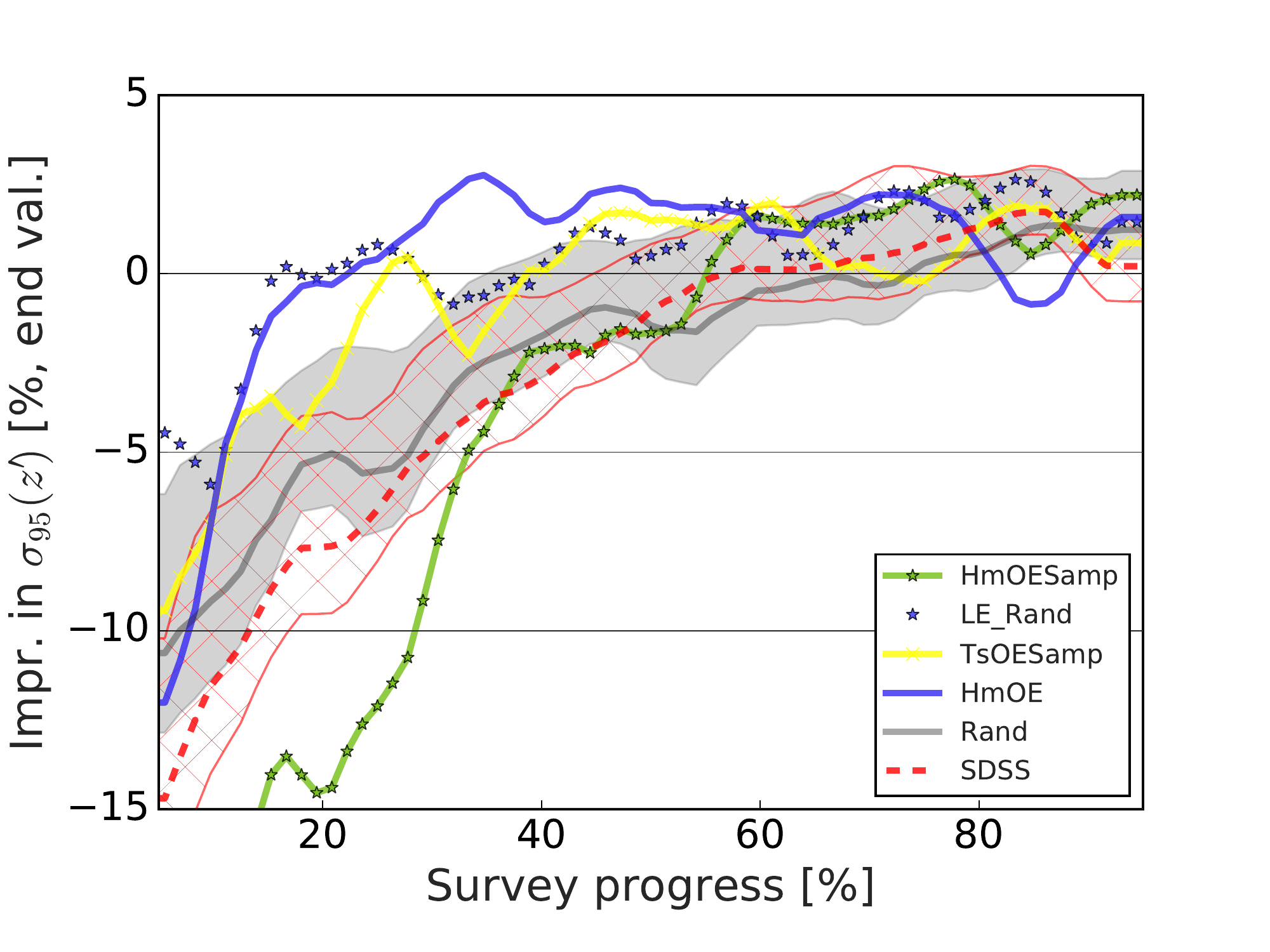}
   \includegraphics[scale=0.46,clip=true,trim=0 15 40 30]{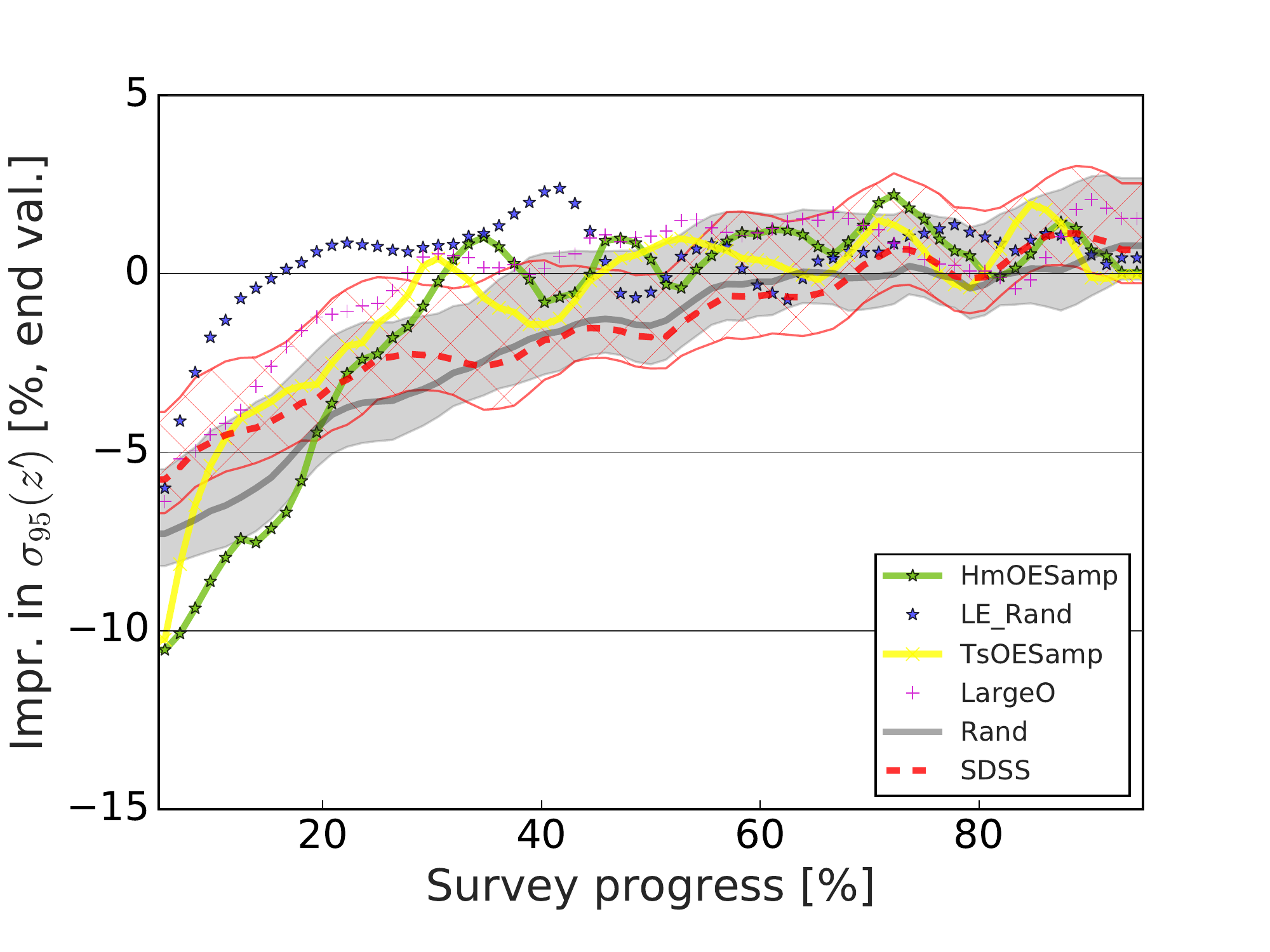}\\
   \includegraphics[scale=0.46,clip=true,trim=0 15 40 30]{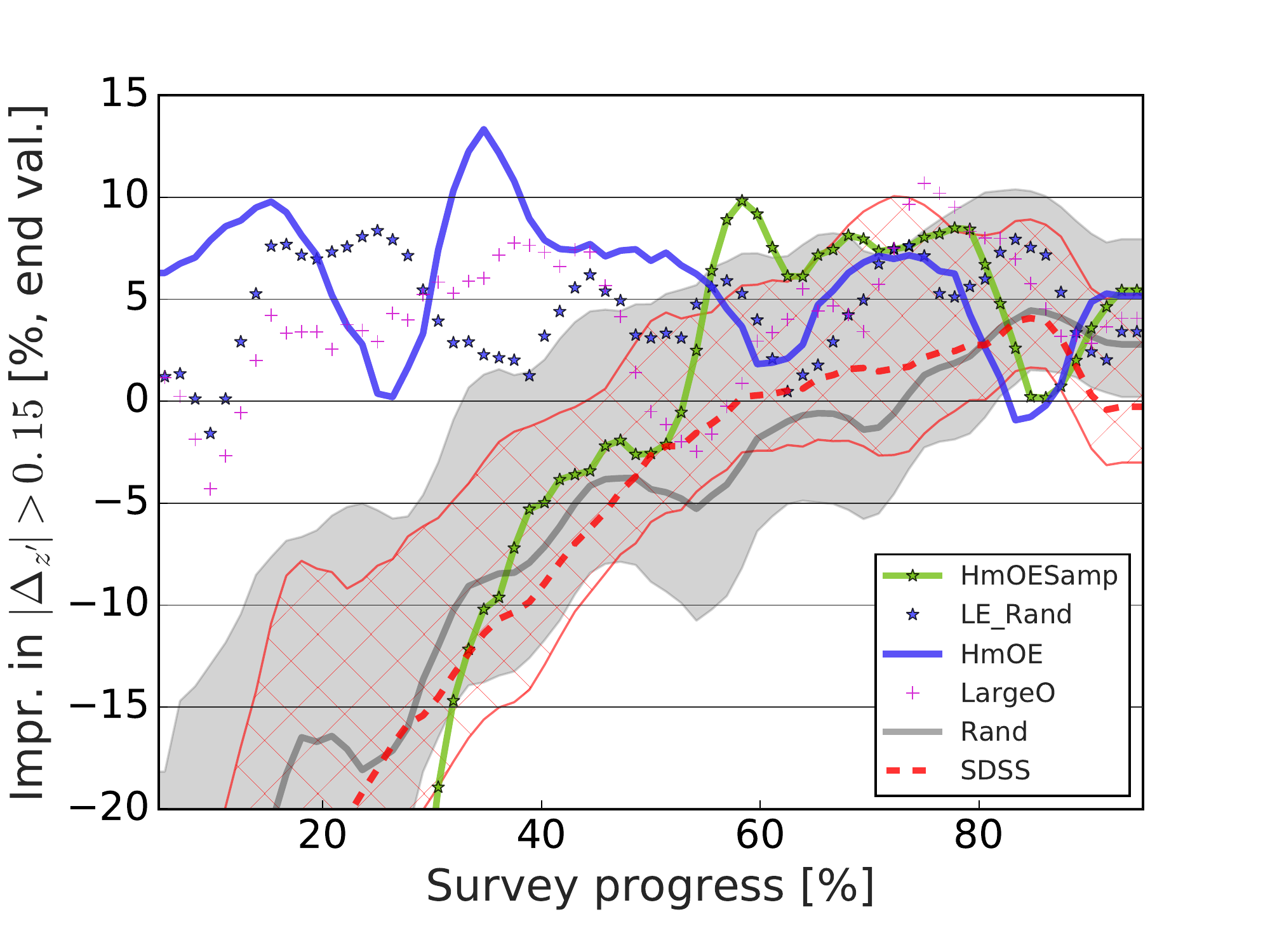}
   \includegraphics[scale=0.46,clip=true,trim=0 15 40 30]{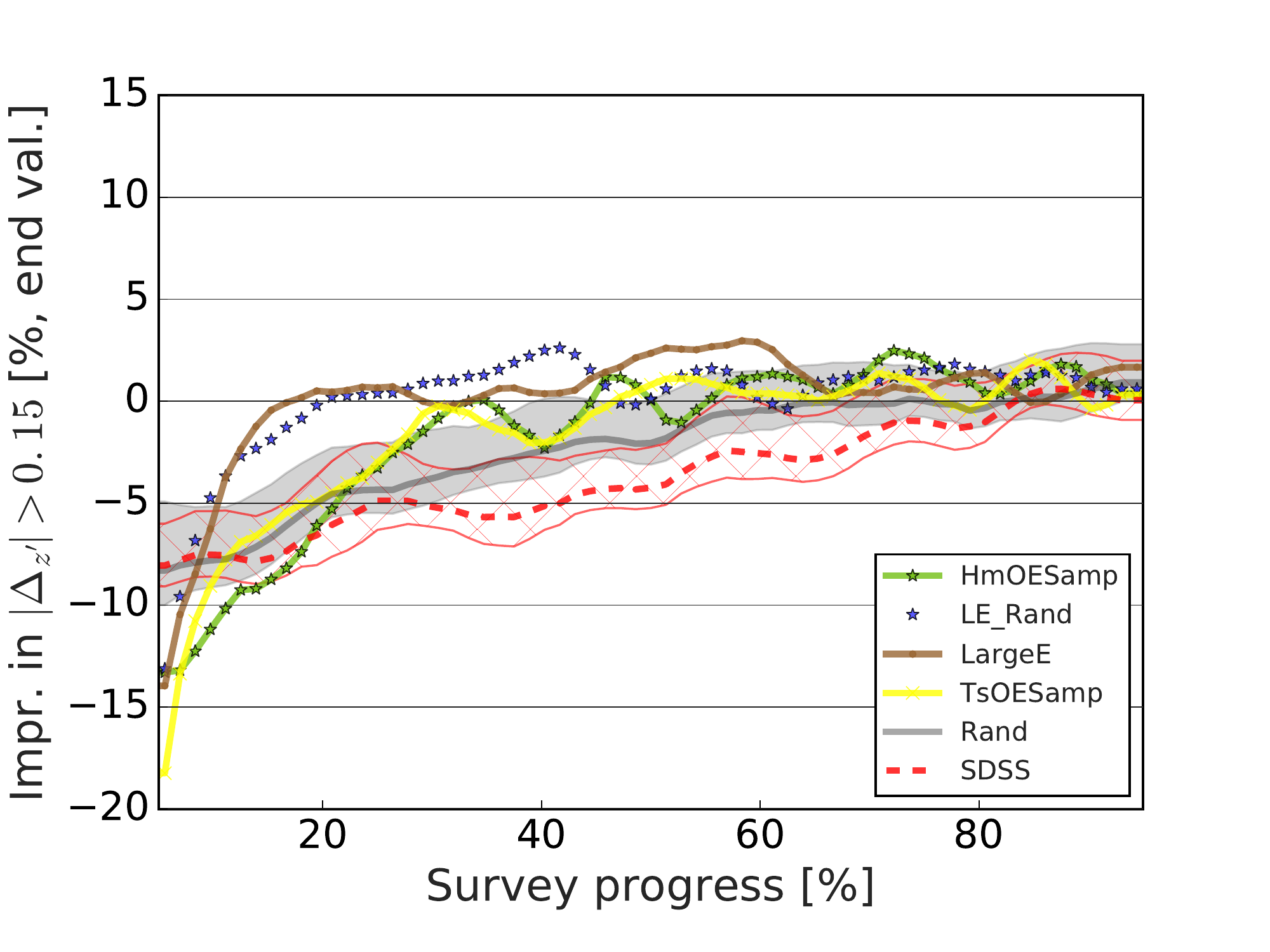}
   \caption{ \label{targetSelectionEffect2} The relative improvement of each measured metric (see y-axis label) value for different targeting algorithms with respect to the final (or end) value from the SDSS date ordered targeting algorithm. The left-hand (right-hand) panels show the results for the SDSS I\&II (SDSS III) analyses. The dark grey shaded region corresponds to the 68\% spread of metric values using the random targeting algorithm (Rand) across {\rr 7} independent experiments, and  {\rrr the red dashed lines and hatched} regions corresponds to the spread of metric values using the SDSS algorithm.} 
\end{figure*}

We next present the same analysis as in the previous subsection, but show the improvement in the measured metrics with respect to the final value obtained at the end of all of the observing runs. We expect all of the routines to converge at the end of the survey, because they have exhaustively selected the target list. This method of presentation allows us to also explore the improvement of metrics using the SDSS targeting algorithm as more training galaxies are added, as shown in Fig. \ref{targetSelectionEffect2}, however we note that this result has been present before \citep[e.g.][]{2014arXiv1410.4696H}, within the context of randomly selecting training data. The {\rrr red dashed lines and hatched region} correspond to the median and 68\% dispersion for each metric as measured using the galaxies targeted using the SDSS algorithm. The left-hand panels again show the results of the SDSS I\&II analysis and the right-hand panels show the results of the SDSS III analysis.

Examining the top panels of Fig. \ref{targetSelectionEffect2} we find that all of the algorithms require at least 50\% of the available data before the measured values of $\sigma_{68}(z')$ approach the final value as measured by the SDSS. We note that the values at the right edge of many of the panels are slightly above the 0\% improvement mark, which is due to the smoothing and dispersion of the results which causes the final value of the metrics to be biased slightly low. We find that the hybrid targeting algorithm LE\_Rand again performs well compared to both the random (Rand) and the time ordered SDSS targeting algorithms, and some of the other targeting algorithms, for example, LargeE, TsBE and LargeB all perform substantially worse than the random algorithm until around 50\% of the runs have been completed in most metrics.Examining the bottom two rows of Fig. \ref{targetSelectionEffect2} we find that the performance of many of the targeting algorithms, including LE\_Rand, LargeE, and LargeO show an improvement over both the random algorithm, and the SDSS algorithm. In particular after about 20\% of the survey has been completed, these algorithms already produce values which are consistent with the final estimates for both of the metrics $\sigma_{95}(z')$ and the outlier fraction $|\Delta_{z'}|>0.15$. This can be compared to the amount of the survey  60\% (80\%) which is required by the SDSS and random algorithms  for the SDSS I\&II (III) targeted galaxies before obtaining metric values which are consistent with the final values.

{\rr The values of the measured statistics approach the final values after approximately 40-60\% of the simulated survey runs, depending on the target selection algorithm. This suggests that the information content of the training set is already saturated for this fixed target pool. In the next subsection we allow a much enlarged target pool to be drawn from. The motivation for this is that one could potentially change survey strategies once the redshift estimates stabilise.}

%

\subsubsection{Enlarged potential candidate list}
\label{moredata}
Rather than being constrained to the same samples of target candidates as in \S \ref{inst} \& \S\ref{final}, we now allow the machine learning targeting algorithms to select targets from a much larger target pool. This could be realised in practise by noting that some of the algorithms, such as LE\_Rand produce good redshift estimates within the first 40\% or 50\% of the available survey time. One could then change follow-up survey strategies and construct a new target list drawn from fainter galaxies, which require longer observing times or resources, or new target lists from an enlarged survey footprint.

To explore this approach we generate two training datasets labelled `tr1' and `tr2', and one test data sample in each of the separate SDSS I\&II and SDSS III analysis. The sizes of tr1 and the test sample are fixed to 100k. The size and selection of tr2 corresponds to all galaxies which are not in the test sample and is of size 770k for the SDSS I\&II analysis and 1011k for the SDSS III analysis. We note that this implies that tr1 is a subsample of tr2, and both tr1 and tr2 are independent from the test samples.

\begin{figure*}
   \centering
   \includegraphics[scale=0.46,clip=true,trim=0 15 40 30]{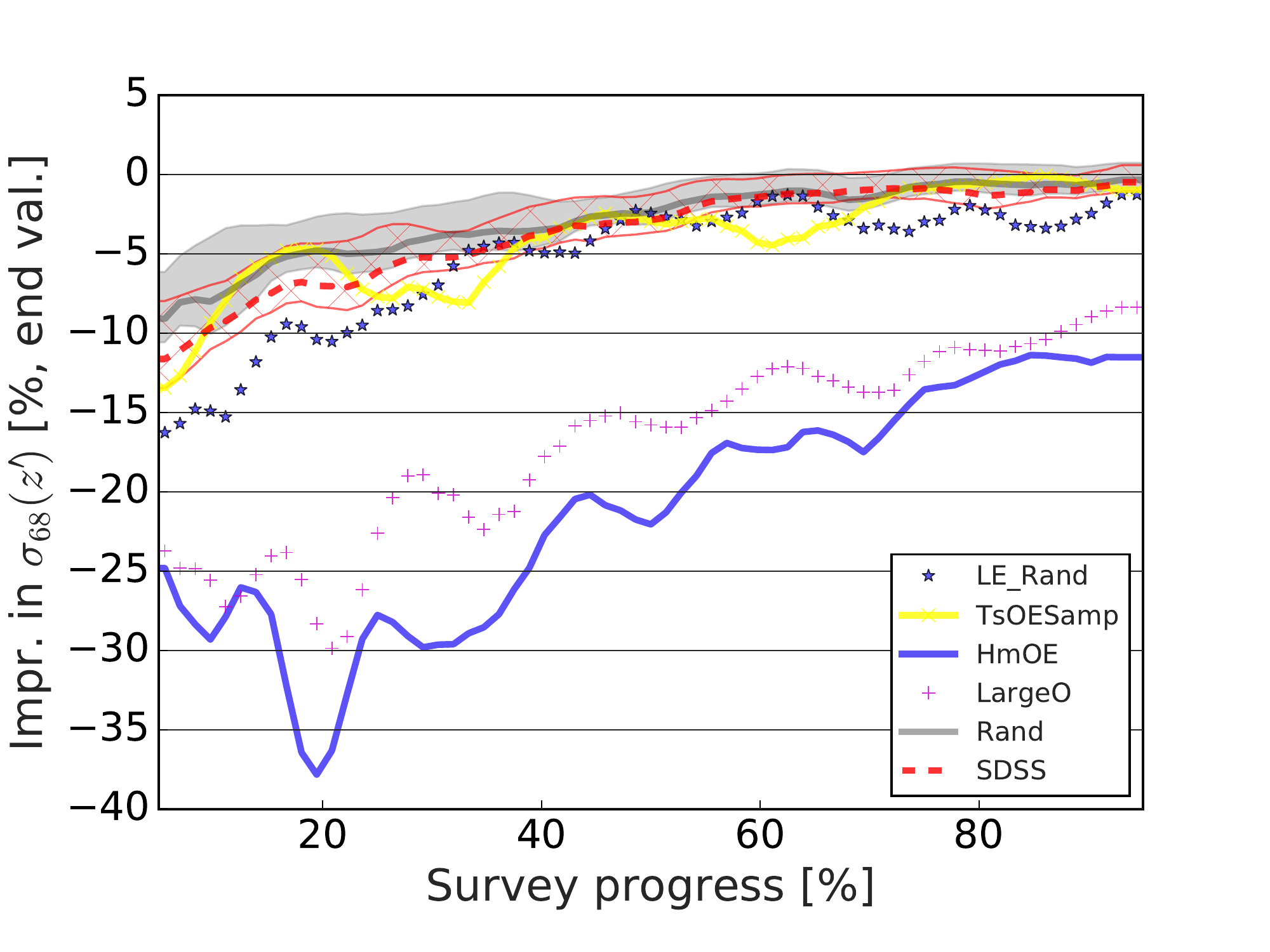}
   \includegraphics[scale=0.46,clip=true,trim=0 15 40 30]{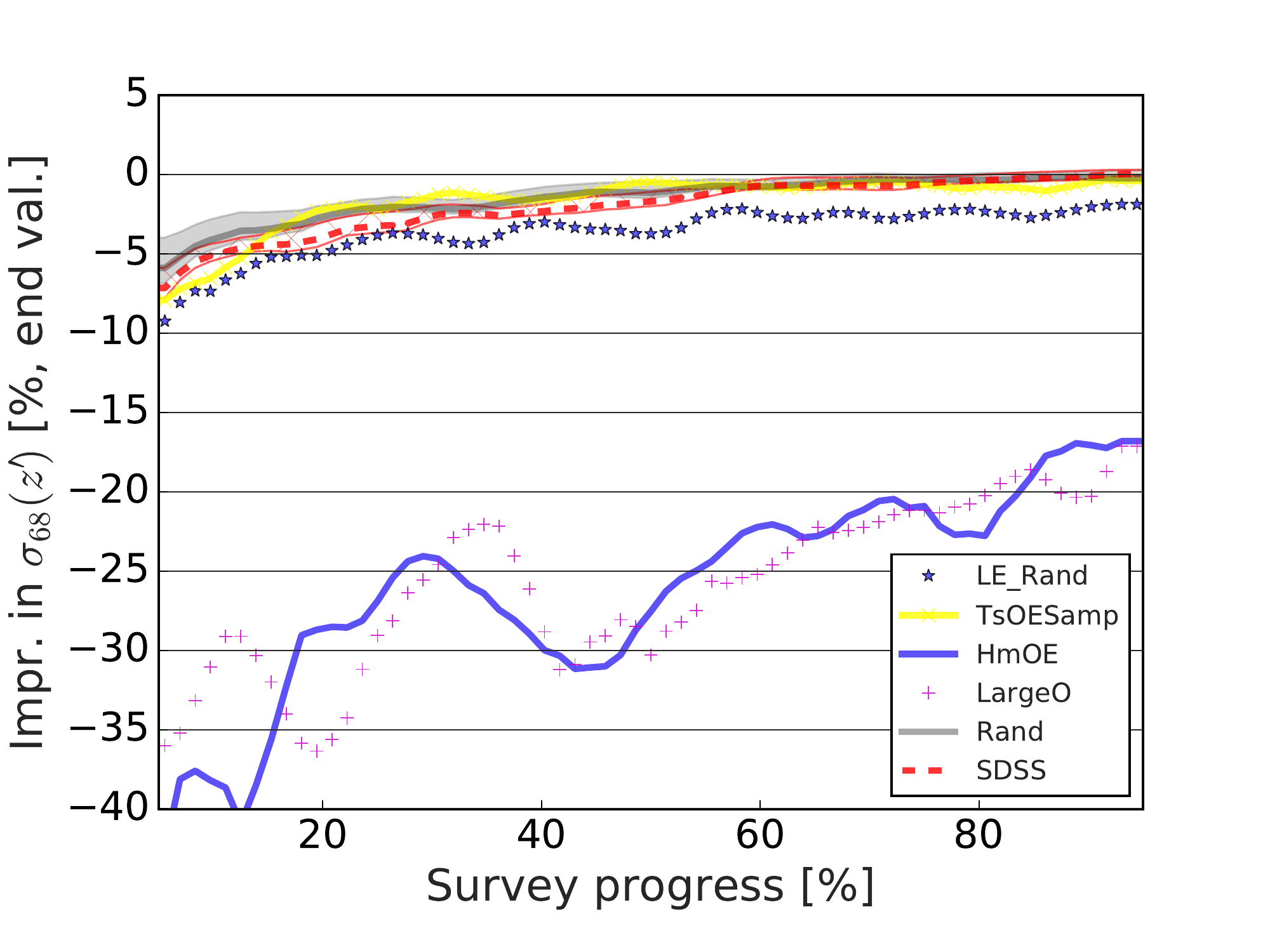}\\
   \includegraphics[scale=0.46,clip=true,trim=0 15 40 30]{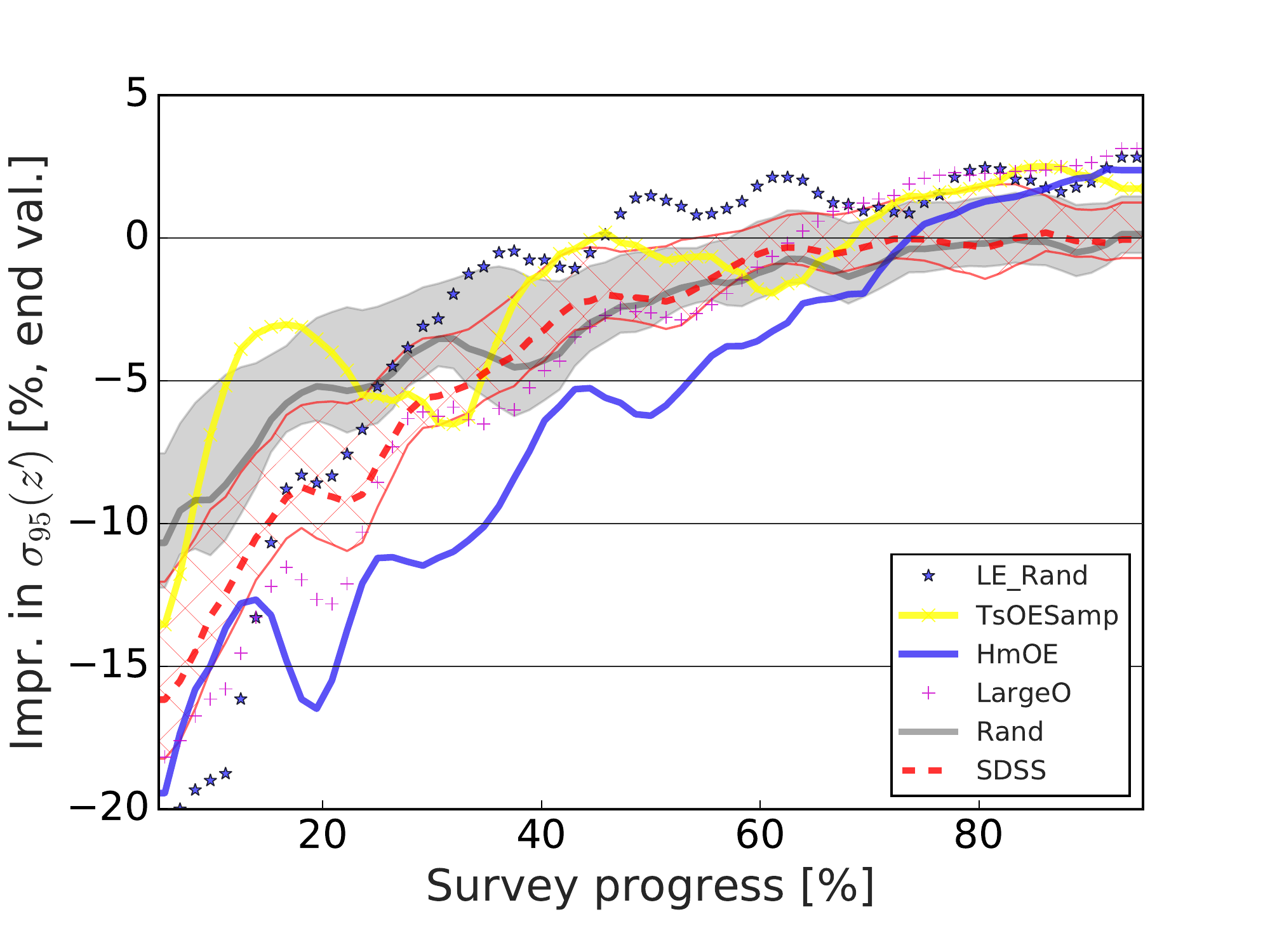}
   \includegraphics[scale=0.46,clip=true,trim=0 15 40 30]{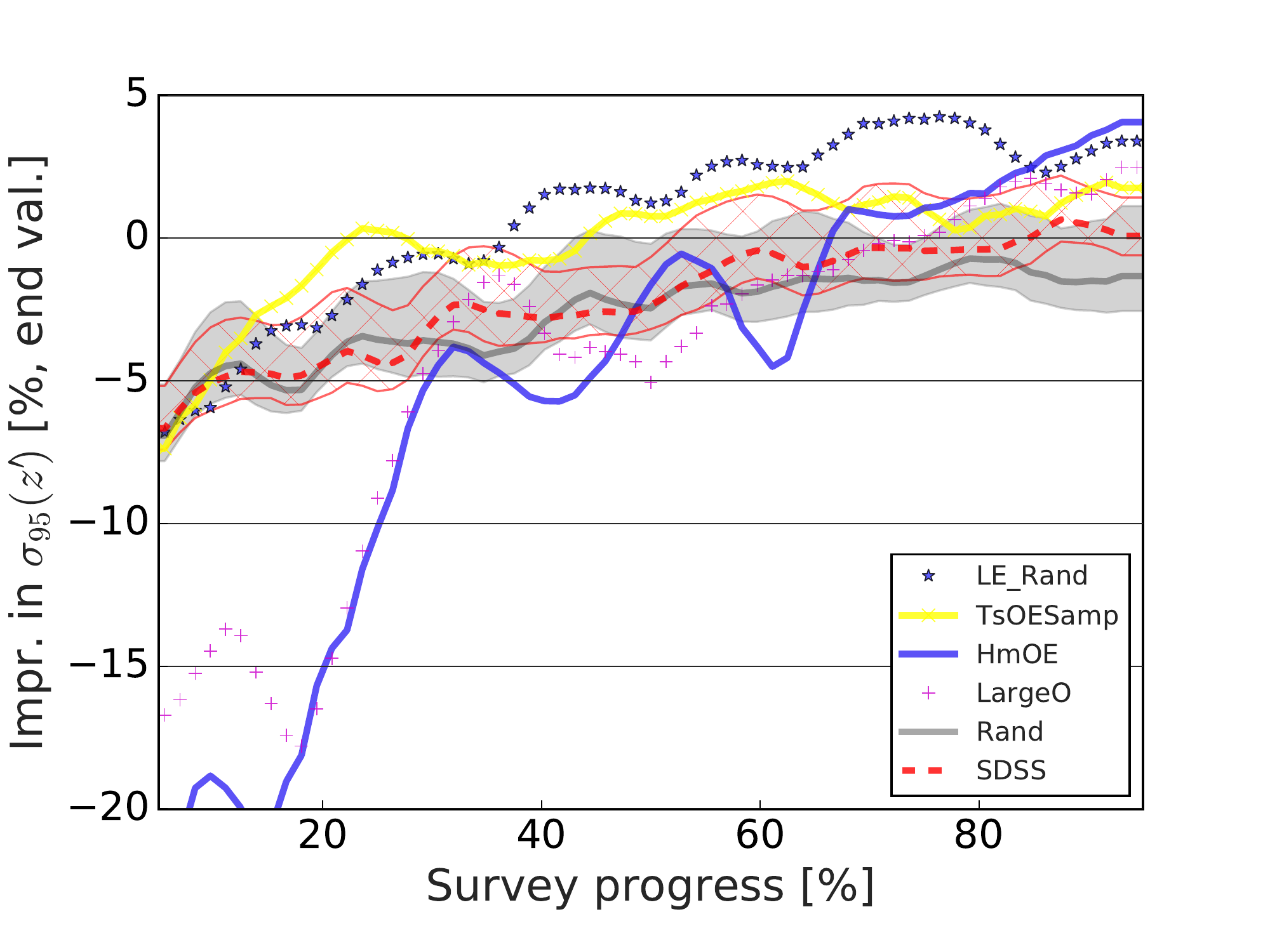}\\
   \includegraphics[scale=0.46,clip=true,trim=0 15 40 30]{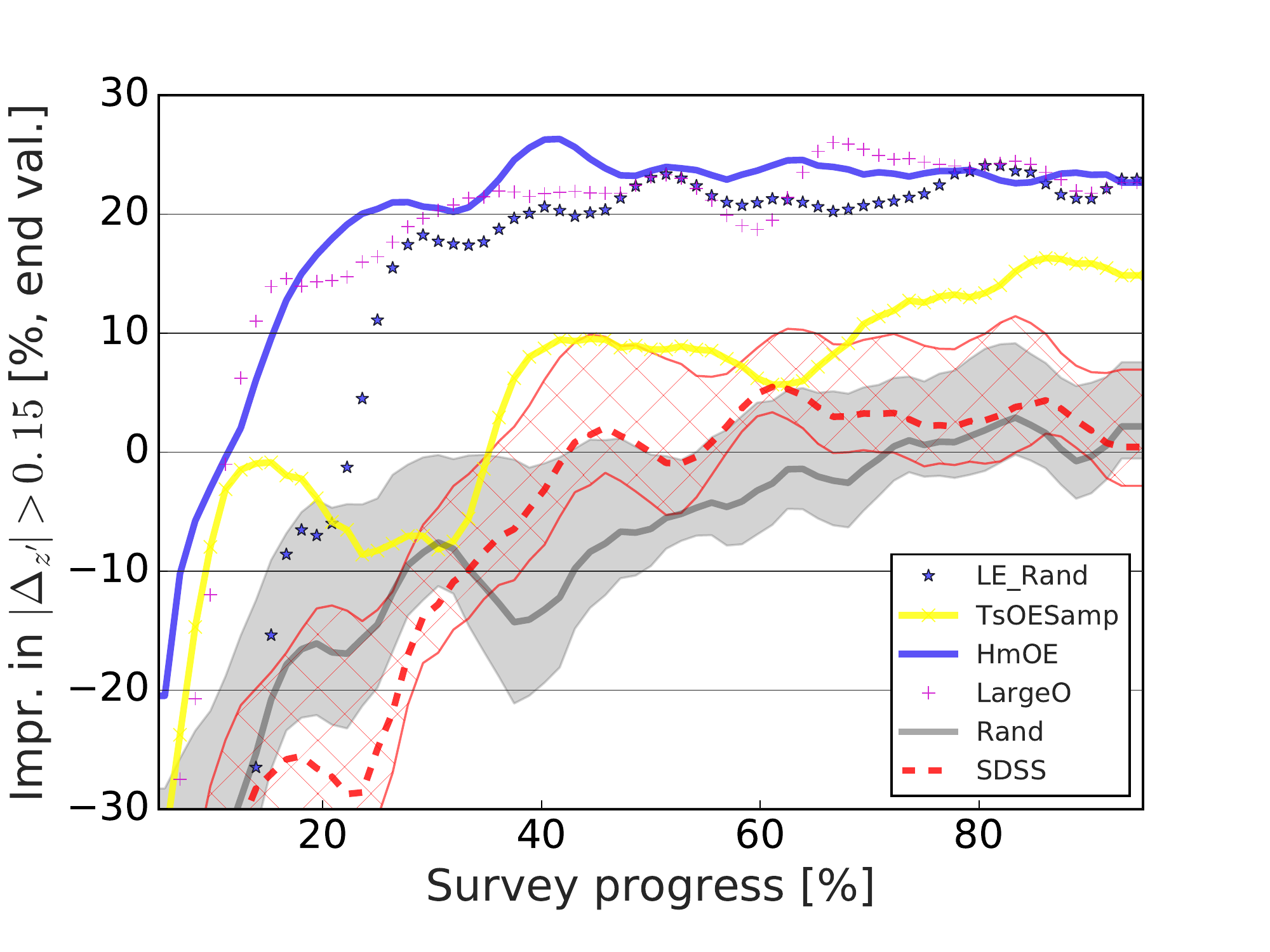}
   \includegraphics[scale=0.46,clip=true,trim=0 15 40 30]{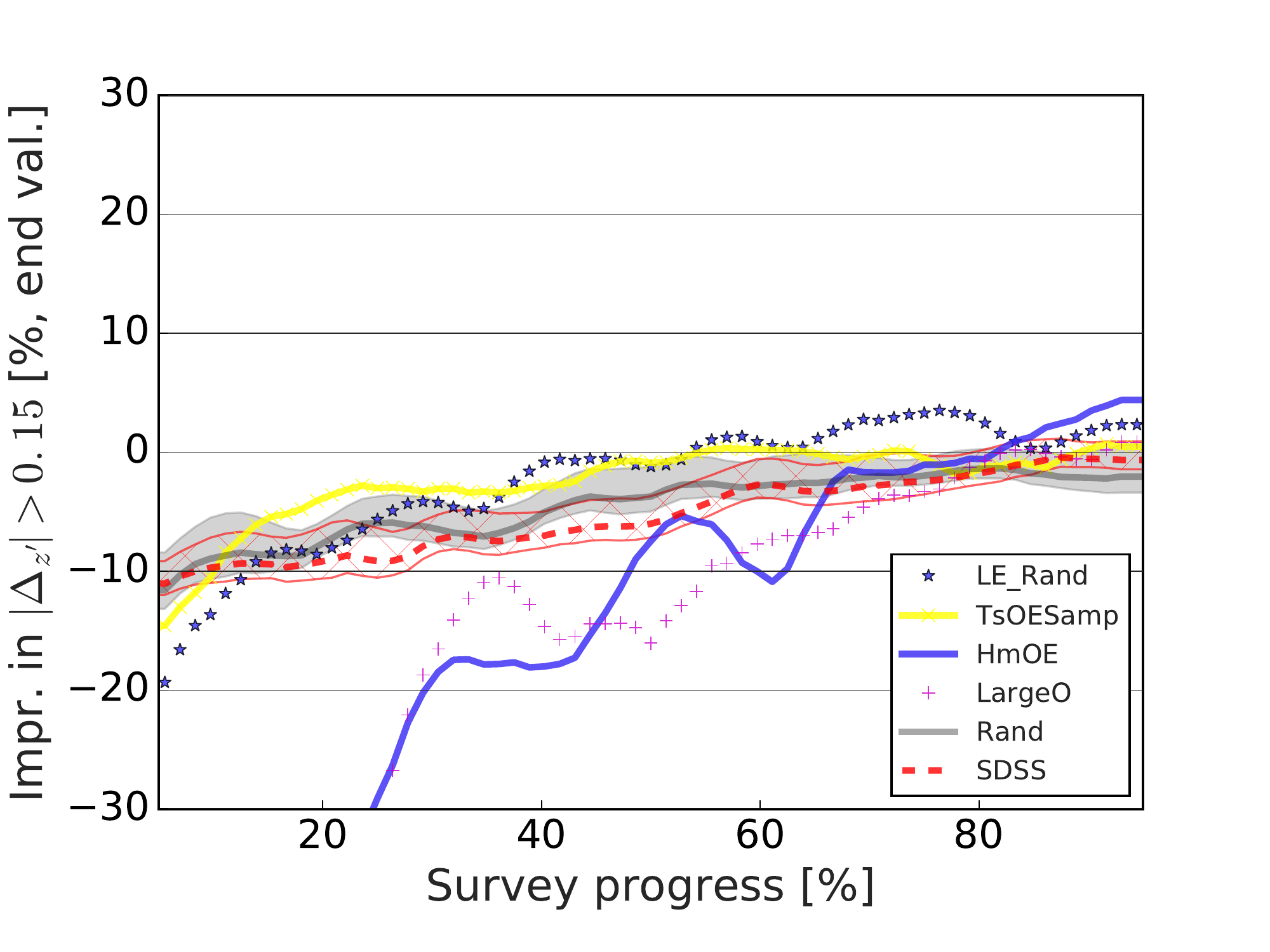}\\
   \caption{ \label{targetSelectionEffect6} The relative improvement of each measured metric (see the y-axis labels) value for different targeting algorithms with respect to the final (or end) value from the SDSS date ordered targeting algorithm. The left-hand (right-hand) panels show the results for the SDSS I\&II (SDSS III) analyses. The grey shaded region corresponds to the 68\% spread of metric values using the random targeting algorithm (Rand) across {\rr 7}  independent experiments, and {\rrr the red dashed lines and hatched}  regions corresponds to the spread of metric values using the date ordered SDSS targeting algorithm. In this figure only, the machine learning targeting algorithms may draw targets from a much larger target pool than the SDSS and Rand algorithms.} 
\end{figure*}

We restrict the random and SDSS targeting algorithms to the smaller data set tr1, but allow the other targeting algorithms the freedom to select from the larger pool of potential target galaxies tr2. In Fig. \ref{targetSelectionEffect6} we present the relative improvement of each metric (see the y-axis labels) value for different targeting algorithms with respect to the final value from the SDSS time ordered targeting algorithm.

By examining all panels of Fig. \ref{targetSelectionEffect6} we again find that after approximately 20\% of the observing runs have been taken, the hybrid target selection algorithm (LE\_Rand) performs as least as well and often better on all metrics in both sets of analysis than both the random algorithm, and the SDSS algorithm. We remind the reader that the SDSS algorithm was not designed to be optimised for machine learning redshift estimation, and therefore a direct comparison is included here for completeness.  We find that the two other target selection algorithms LargeB and HmBE perform poorly on the metrics $\sigma_{68}(z')$, less poorly on the metric $\sigma_{95}(z')$, and perform very well on the outlier fraction metric $|\Delta_{z'}|>0.15$. This suggests that these algorithms are targeting galaxies which remove the outliers from the tails of the distribution, but are not improving the redshift estimates of the majority of the test sample galaxies. We note that the improvement of the hybrid algorithm LE\_Rand is much less pronounced for the SDSS III sample than for the SDSS I\&II sample, which is probably due to the more homogeneous sample of SDSS III galaxies.

{\rr One can directly see the effect of allowing an enlarged target pool by examining both Figs \ref{targetSelectionEffect2} \& \ref{targetSelectionEffect6}. The random and SDSS lines and contours are equivalent in both figures, and the improvement in the metric values of some of the machine learning target selection algorithms is noticeable.}

%% file: conclus.tex
Photometric galaxy catalogues can be maximally exploited for cosmological analyses once galaxy redshifts have been measured spectroscopically, or estimated photometrically. Machine learning architectures can map measured photometric features onto spectroscopic redshifts for a subset of `training' data with both quantities. This mapping can then be used to estimate redshifts for all photometrically selected galaxies which are representative of the training sample. Constructing a training set often requires programs of dedicated telescope time to perform spectroscopic follow up of a set of target galaxies for which one may reasonably assume a redshift can be measured. In this paper we propose a machine learning based target selection algorithm which leads to faster improvements in the machine learning redshift estimation of a final sample of test galaxies, such that less spectra are required.

We explore {\rr 7 different target selection algorithms, in addition to the SDSS, and a random target selection algorithm} which are constructed by applying machine learning techniques in a subtly different way than normal redshift estimation. Instead of estimating the value of the photometric redshift we construct a machine to predicted the size of the photometric redshift error and the photometric offset, or bias, defined as $|z_{spec}-z_{ML}|$. We find that these predictions are very accurate, for example $(2\pm5)\times10^{-3}$ for photometric redshift errors, which, for SDSS data,  typically have values greater than $2\times10^{-2}$. We combine these predictions to construct {\rr 7 different machine learning target selection algorithms}. We find that a hybrid algorithm which selects equal numbers of random galaxies (denoted by `Rand' in Table. \ref{TargetSelectionCheatSheet}), and galaxies with the largest predicted redshift error (LargeE) provides the best target selection routine of those examined. 

To test our methods we construct sets of experiments using data drawn from the Sloan Digital Sky Survey Data Release 12 \citep[SDSS DR12][]{2015arXiv150100963A}. We perform two independent sets of analyses by selecting data observed within the time frames of SDSS I\&II and SDSS III. For the purposes of this work, we construct clean samples of galaxies, free of stellar contamination. {\rr One} may also choose to select combinations of stars, galaxies, and other artefacts if the science goal were to improve star/galaxy separation using machine learning.

To compare the different target selection methods we compute photometric redshifts on an independent sample of representative galaxies and measure the following metrics:  $\sigma_{68}(z')$ which is defined to be the 68\% dispersion of the redshift scaled residuals $\Delta_{z'}=(z_{spec}-z_{ML})/(1+z_{spec})$, and would be equivalent to 1 standard deviation if  $\Delta_{z'}$ were well described by a Gaussian; and the values of the 95\% dispersion of $\Delta_{z'}$ defined as  $\sigma_{95}(z')$; as well as the outlier fraction defined by the fraction of data for which $|\Delta_{z'}|>0.15$.

We compare the {\rr 7 } targeting algorithms presented in this work with a random target selection algorithm, and with the SDSS targeting order as defined by the date when the spectra were obtained. We find that after 20\% of simulated runs, the hybrid targeting algorithm performs no worse than the random and SDSS targeting algorithms on the metric $\sigma_{68}(z')$. We find that after only 10\% of the total number of observing runs the value of $\sigma_{95}(z')$, and the outlier fraction defined are between 5\% and 20\% better than {\rr both a random selection algorithm, and } that of the SDSS. We note that the SDSS target selection algorithms \citep[see ][]{2000AJ....120.1579Y,2001AJ....122.2267E,2002AJ....123.2945R} were not constructed for optimal machine learning redshift estimation, and therefore a direct comparison between the SDSS and the targeting algorithms described in this work is for presentation purposes only. It however shows that machine learning based target selection algorithms can be used to tailor spectroscopic follow-up for specific science goals.

We also estimate how many observing runs are required before the metric values, which are measured on the independent test samples, approach the final metric values, as calculated using all of the target galaxies. We find that the hybrid method is able to reach the same precision in the outlier fraction and $\sigma_{95}(z')$ within 20\% of the total amount of followed-up targets. To obtain similar precision on the value of $\sigma_{68}(z')$ one requires between 40\% (60\%) of observed targets for the SDSS I\&II (III) data sample.

Assuming that one may reach the same precision with less observing runs, suggests that this method could be used to change observing strategies for the latter part of a follow up survey, in order to target different samples of galaxies. This could be used to increase the footprint of the survey, which could help with, for example, clustering redshift estimates \citep[see][]{2013arXiv1303.4722M}, or to probe to deeper apparent magnitude limits which require longer integration times. Of course such a target selection algorithm would also have to be convolved with the science goals of the survey. For example for Baryon Acoustic Oscillation surveys, one requires large area samples of homogeneous galaxies.

We estimate how a difference in the selected targets could effect the recovered metrics, by allowing all of the {\rr  machine learning targeting routines, i.e. except} the time ordered SDSS and random algorithms, to draw targets from a much enlarged pool of target galaxies. Using the hybrid targeting algorithm we find that the outlier fraction is improved by 20\% compared to that of the SDSS or random algorithms. A further extension of this work would be to change the test sample, as the target sample becomes enlarged, allowing the test sample to cover a larger range of photometric features values. This will be explored in future work.

During this work we have made a few simplifying assumptions. For example we have assumed that each potential target could be observed in the run of our choice. {\rr In practice this may not be possible due to observing conditions such as the galactic coordinates of each target, or due to fiber collisions for targets which are nearby on the sky.} Instead one could prepare a list of potential targets which could be viewed in the subsequent target run, and then determine which of these targets would be the best to follow up. Furthermore most surveys are not designed to have a purely photometric phase which would identify all target candidates, followed by a dedicated spectroscopic follow up. In these cases the techniques presented in this paper could still be applied, by introducing and varying an ever increasing set of potential targets as more photographic data is obtained. Should one wish to use this technique for surveys that need to define a spectroscopic target selection before initial imaging or spectra have been taken, one may augment the dataset using existing imaging of comparable depth, or using realistic simulations, and perform a similar analysis described in this work. 

{\rrr One import point to note is that in this work we have assumed that the sample of galaxies with spectra is representative of those galaxies without spectra, and that test sample spectra may be obtained. This is not necessarily the case in reality due to, for example, the difficulty of obtaining spectra for very faint galaxies or objects which sit in the redshift desert. A future approach to explore this problem could be to bias the initial training sample by applying sets of color cuts and repeating this analysis. Then we would have to further impose that the selected sample for spectroscopic follow-up remains biased in some sense (perhaps in brightness or color) with respect to a final test sample. The exact nature of the biases we should incur would vary for each observation survey.}

{\rr Other extensions could be to combine the estimation of both the bias, $b_{z_{ML}}$ and error $\delta_{z_{ML}}$ simultaneously, not separately, as performed here (see \S\ref{prediction}). Many machine learning algorithms, including decision tree based methods allow such a multiple output feature training.} Furthermore we have not explored the use of other machine learning algorithms to perform the prediction or redshift estimation tasks. One could extend this work by  exploring different algorithms. We choose to use random forests because they are easily run in parallel across many cores, and they are inherently very fast because they use decision trees. Random forests are also very robust machine learning frameworks which perform well compared with other methods on many problems in the literature \citep[e.g.,][]{2010A&A...523A..31H,2014arXiv1406.4407S}. One could also extend this analysis by comparing the full redshift distribution of the photometric and spectroscopic samples, instead of the metrics chosen in this paper.

The target selection methods presented in this paper could be applied to the Dark Energy Survey \citep{2005astro.ph.10346T} or Euclid \citep{2011arXiv1110.3193L} by first defining a training sample and using this to predict which target galaxies will have large redshift errors and biases in a machine learning approach, and then iteratively following up sub samples of galaxies and retrain the prediction models. It could also be applied with other science goals in mind, for example to star and galaxy separation, or to identify samples of galaxies within a particular population class, or redshift range (for example see \footnote{sagasurvey.org}).

\begin{appendices}
\section{CasJobs MySQL query}
We obtain observational data from the SDSS using the following MySQL query which is run in the DR12 schema:
\label{sdss_q1}
\begin{verbatim}
select s.specObjID, q.objid,  
q.dered_u, q.dered_g, q.dered_r, q.dered_i, q.dered_z, 
q.modelMagErr_u, q.modelMagErr_g, q.modelMagErr_r, 
q.modelMagErr_i, q.modelMagErr_z, 
q.petroRad_r,
q.type as photpType, 
s.z as specz, s.zerr as specz_err,
s.zWarning, s.mjd
into mydb.specPhotoDR12v3 from SpecObjAll as s 
join photoObjAll as q on s.bestobjid=q.objid 
\end{verbatim}
\end{appendices}
This produces 3.9 million results of which we further process as described in \S\ref{obs_data}.

%% file: target_selection.bbl
\begin{thebibliography}{}

\bibitem[\protect\citeauthoryear{{Alam}, {Albareti} \& et al.}{{Alam}
  et~al.}{2015}]{2015arXiv150100963A}
{Alam} S.,  {Albareti} F.~D.,    et al. 2015, ArXiv e-prints:1501.00963

\bibitem[\protect\citeauthoryear{{Bonnett}}{{Bonnett}}{2015}]{2013arXiv1312.1287B}
{Bonnett} C.,  2015, \mnras, 449, 1043

\bibitem[\protect\citeauthoryear{Breiman}{Breiman}{2001}]{RandoMforests}
Breiman L.,  2001, Machine Learning, 45, 5

\bibitem[\protect\citeauthoryear{Breiman, Friedman, Olshen \& Stone}{Breiman
  et~al.}{1984}]{ig}
Breiman L.,  Friedman J.~H.,  Olshen R.~A.,    Stone C.~J.,  1984,
  Classification and Regression Trees.
Wadsworth International Group, Belmont, CA

\bibitem[\protect\citeauthoryear{{Carliles}, {Budav{\'a}ri}, {Heinis}, {Priebe}
  \& {Szalay}}{{Carliles} et~al.}{2008}]{2008ASPC..394..521C}
{Carliles} S.,  {Budav{\'a}ri} T.,  {Heinis} S.,  {Priebe} C.,    {Szalay} A.,
  2008, in {Argyle} R.~W.,  {Bunclark} P.~S.,   {Lewis} J.~R.,  eds,
  Astronomical Data Analysis Software and Systems XVII Vol.~394 of Astronomical
  Society of the Pacific Conference Series, {Photometric Redshift Estimation on
  SDSS Data Using Random Forests}.
p.~521

\bibitem[\protect\citeauthoryear{{Carrasco Kind} \& {Brunner}}{{Carrasco Kind}
  \& {Brunner}}{2013}]{tpz}
{Carrasco Kind} M.,  {Brunner} R.~J.,  2013, \mnras, 432, 1483

\bibitem[\protect\citeauthoryear{{Collister} \& {Lahav}}{{Collister} \&
  {Lahav}}{2004}]{2004PASP..116..345C}
{Collister} A.~A.,  {Lahav} O.,  2004, \pasp, 116, 345

\bibitem[\protect\citeauthoryear{{Csabai}, {Dobos}, {Trencs{\'e}ni},
  {Herczegh}, {J{\'o}zsa}, {Purger}, {Budav{\'a}ri} \& {Szalay}}{{Csabai}
  et~al.}{2007}]{2007AN....328..852C}
{Csabai} I.,  {Dobos} L.,  {Trencs{\'e}ni} M.,  {Herczegh} G.,  {J{\'o}zsa} P.,
   {Purger} N.,  {Budav{\'a}ri} T.,    {Szalay} A.~S.,  2007, Astronomische
  Nachrichten, 328, 852

\bibitem[\protect\citeauthoryear{{Dieleman}, {Willett} \& {Dambre}}{{Dieleman}
  et~al.}{2015}]{2015arXiv150307077D}
{Dieleman} S.,  {Willett} K.~W.,    {Dambre} J.,  2015, ArXiv: 1503.07077

\bibitem[\protect\citeauthoryear{{Eisenstein}, {Annis}, {Gunn}, {Szalay},
  {Connolly}, {Nichol} et~al.,}{{Eisenstein}
  et~al.}{2001}]{2001AJ....122.2267E}
{Eisenstein} D.~J.,  {Annis} J.,  {Gunn} J.~E.,  {Szalay} A.~S.,  {Connolly}
  A.~J.,  {Nichol} R.~C.,    et~al., 2001, \aj, 122, 2267

\bibitem[\protect\citeauthoryear{{Eisenstein}
  D.~J.}{{Eisenstein}}{2011}]{2011AJ....142...72E}
{Eisenstein} D.~J. e.~a.,  2011, \aj, 142, 72

\bibitem[\protect\citeauthoryear{{Gerdes}, {Sypniewski}, {McKay}, {Hao},
  {Weis}, {Wechsler} \& {Busha}}{{Gerdes} et~al.}{2010}]{2010ApJ...715..823G}
{Gerdes} D.~W.,  {Sypniewski} A.~J.,  {McKay} T.~A.,  {Hao} J.,  {Weis} M.~R.,
  {Wechsler} R.~H.,    {Busha} M.~T.,  2010, \apj, 715, 823

\bibitem[\protect\citeauthoryear{{Gunn}, {Siegmund}, {Mannery}, {Owen}, {Hull},
  {Leger}, {Carey}, {Knapp}, {York}, {Boroski}, {Kent}, {Lupton}, {Rockosi}
  et~al.,}{{Gunn} et~al.}{2006}]{Gunn:2006tw}
{Gunn} J.~E.,  {Siegmund} W.~A.,  {Mannery} E.~J.,  {Owen} R.~E.,  {Hull}
  C.~L.,  {Leger} R.~F.,  {Carey} L.~N.,  {Knapp} G.~R.,  {York} D.~G.,
  {Boroski} W.~N.,  {Kent} S.~M.,  {Lupton} R.~H.,  {Rockosi} C.~M.,    et~al.,
  2006, \aj, 131, 2332

\bibitem[\protect\citeauthoryear{{H{\'a}la}}{{H{\'a}la}}{2014}]{2014arXiv1412.8341H}
{H{\'a}la} P.,  2014, ArXiv e-prints:1412.8341

\bibitem[\protect\citeauthoryear{Hastie, Tibshirani \& Friedman}{Hastie
  et~al.}{2001}]{hastie01statisticallearning}
Hastie T.,  Tibshirani R.,    Friedman J.,  2001, The Elements of Statistical
  Learning.
Springer Series in Statistics, Springer New York Inc., New York, NY, USA

\bibitem[\protect\citeauthoryear{{Hildebrandt}, {Arnouts}, {Capak},
  {Moustakas}, {Wolf} \& {Abdalla}}{{Hildebrandt}
  et~al.}{2010}]{2010A&A...523A..31H}
{Hildebrandt} H.,  {Arnouts} S.,  {Capak} P.,  {Moustakas} L.~A.,  {Wolf} C.,
   {Abdalla} e.~a.,  2010, \aap, 523, A31

\bibitem[\protect\citeauthoryear{{Hogan}, {Fairbairn} \& {Seeburn}}{{Hogan}
  et~al.}{2015}]{2015MNRAS.449.2040H}
{Hogan} R.,  {Fairbairn} M.,    {Seeburn} N.,  2015, \mnras, 449, 2040

\bibitem[\protect\citeauthoryear{{Hoyle}}{{Hoyle}}{2015}]{2015arXiv150407255H}
{Hoyle} B.,  2015, ArXiv:1504.07255

\bibitem[\protect\citeauthoryear{{Hoyle}, {Rau}, {Bonnett}, {Seitz} \&
  {Weller}}{{Hoyle} et~al.}{2015}]{2015arXiv150106759H}
{Hoyle} B.,  {Rau} M.~M.,  {Bonnett} C.,  {Seitz} S.,    {Weller} J.,  2015,
  \mnras, 450, 305

\bibitem[\protect\citeauthoryear{{Hoyle}, {Rau}, {Paech}, {Bonnett}, {Seitz} \&
  {Weller}}{{Hoyle} et~al.}{2015}]{2015arXiv150308214H}
{Hoyle} B.,  {Rau} M.~M.,  {Paech} K.,  {Bonnett} C.,  {Seitz} S.,    {Weller}
  J.,  2015, \mnras, 452, 4183

\bibitem[\protect\citeauthoryear{{Hoyle}, {Rau}, {Zitlau}, {Seitz} \&
  {Weller}}{{Hoyle} et~al.}{2015}]{2014arXiv1410.4696H}
{Hoyle} B.,  {Rau} M.~M.,  {Zitlau} R.,  {Seitz} S.,    {Weller} J.,  2015,
  \mnras, 449, 1275

\bibitem[\protect\citeauthoryear{{Jouvel}, {Abdalla}, {Kirk}, {Lahav}, {Lin},
  {Annis}, {Kron} \& {Frieman}}{{Jouvel} et~al.}{2014}]{2014MNRAS.438.2218J}
{Jouvel} S.,  {Abdalla} F.~B.,  {Kirk} D.,  {Lahav} O.,  {Lin} H.,  {Annis} J.,
   {Kron} R.,    {Frieman} J.~A.,  2014, \mnras, 438, 2218

\bibitem[\protect\citeauthoryear{Kohonen}{Kohonen}{1997}]{Kohonen:1997:SM:261082}
Kohonen T.,  ed. 1997, Self-organizing Maps.
Springer-Verlag New York, Inc., Secaucus, NJ, USA

\bibitem[\protect\citeauthoryear{{Lahav}}{{Lahav}}{1997}]{1997daa..conf...43L}
{Lahav} O.,  1997, in {Di Gesu} V.,  {Duff} M.~J.~B.,  {Heck} A.,  {Maccarone}
  M.~C.,  {Scarsi} L.,   {Zimmerman} H.~U.,  eds, Data Analysis in Astronomy
  {Artificial neural networks as a tool for galaxy classification.}.
pp 43--51

\bibitem[\protect\citeauthoryear{{Laureijs}, {Amiaux}, {Arduini},
  {Augu{\`e}res}, {Brinchmann}, {Cole} \& et al.}{{Laureijs}
  et~al.}{2011}]{2011arXiv1110.3193L}
{Laureijs} R.,  {Amiaux} J.,  {Arduini} S.,  {Augu{\`e}res} J.~.,  {Brinchmann}
  J.,  {Cole}   et al. 2011, ArXiv e-prints:1110.3193

\bibitem[\protect\citeauthoryear{Li \& Thakar}{Li \&
  Thakar}{2008}]{10.1109/MCSE.2008.6}
Li N.,  Thakar A.~R.,  2008, Computing in Science and Engineering, 10, 18

\bibitem[\protect\citeauthoryear{{Masters}, {Capak}, {Stern}, {Ilbert}
  et~al.,}{{Masters} et~al.}{2015}]{2015ApJ...813...53M}
{Masters} D.,  {Capak} P.,  {Stern} D.,  {Ilbert} O.,    et~al., 2015, \apj,
  813, 53

\bibitem[\protect\citeauthoryear{{Menard}, {Scranton}, {Schmidt}, {Morrison},
  {Jeong}, {Budavari} \& {Rahman}}{{Menard} et~al.}{2013}]{2013arXiv1303.4722M}
{Menard} B.,  {Scranton} R.,  {Schmidt} S.,  {Morrison} C.,  {Jeong} D.,
  {Budavari} T.,    {Rahman} M.,  2013, ArXiv: 1303.4722

\bibitem[\protect\citeauthoryear{{Polsterer}, {Gieseke}, {Igel} \&
  {Goto}}{{Polsterer} et~al.}{2014}]{2014ASPC..485..425P}
{Polsterer} K.~L.,  {Gieseke} F.,  {Igel} C.,    {Goto} T.,  2014, in {Manset}
  N.,  {Forshay} P.,  eds, Astronomical Data Analysis Software and Systems
  XXIII Vol.~485 of Astronomical Society of the Pacific Conference Series,
  {Improving the Performance of Photometric Regression Models via Massive
  Parallel Feature Selection}.
p.~425

\bibitem[\protect\citeauthoryear{{Rau}, {Seitz}, {Brimioulle}, {Frank},
  {Friedrich}, {Gruen} \& {Hoyle}}{{Rau} et~al.}{2015}]{RauEtAllinPrep}
{Rau} M.~M.,  {Seitz} S.,  {Brimioulle} F.,  {Frank} E.,  {Friedrich} O.,
  {Gruen} D.,    {Hoyle} B.,  2015, \mnras, 452, 3710

\bibitem[\protect\citeauthoryear{{Richards}, {Fan}, {Newberg}, {Strauss},
  {Vanden Berk} et~al.,}{{Richards} et~al.}{2002}]{2002AJ....123.2945R}
{Richards} G.~T.,  {Fan} X.,  {Newberg} H.~J.,  {Strauss} M.~A.,  {Vanden Berk}
  D.~E.,    et~al., 2002, \aj, 123, 2945

\bibitem[\protect\citeauthoryear{{S{\'a}nchez}, {Carrasco Kind}, {Lin},
  {Miquel}, {Abdalla} et~al.,}{{S{\'a}nchez}
  et~al.}{2014}]{2014arXiv1406.4407S}
{S{\'a}nchez} C.,  {Carrasco Kind} M.,  {Lin} H.,  {Miquel} R.,  {Abdalla}
  F.~B.,    et~al., 2014, \mnras, 445, 1482

\bibitem[\protect\citeauthoryear{Smith et~al.,}{Smith
  et~al.}{2002}]{Smith:2002pca}
Smith J.~A.,  et~al., 2002, \aj, 123, 2121

\bibitem[\protect\citeauthoryear{{Strauss}, {Weinberg}, {Lupton}, {Narayanan},
  {Annis} et~al.,}{{Strauss} et~al.}{2002}]{2002AJ....124.1810S}
{Strauss} M.~A.,  {Weinberg} D.~H.,  {Lupton} R.~H.,  {Narayanan} V.~K.,
  {Annis} J.,    et~al., 2002, \aj, 124, 1810

\bibitem[\protect\citeauthoryear{{Tagliaferri}, {Longo}, {Andreon},
  {Capozziello}, {Donalek} \& {Giordano}}{{Tagliaferri}
  et~al.}{2003}]{2003LNCS.2859..226T}
{Tagliaferri} R.,  {Longo} G.,  {Andreon} S.,  {Capozziello} S.,  {Donalek} C.,
     {Giordano} G.,  2003, Lecture Notes in Computer Science, 2859, 226

\bibitem[\protect\citeauthoryear{{The Dark Energy Survey Collaboration}}{{The
  Dark Energy Survey Collaboration}}{2005}]{2005astro.ph.10346T}
{The Dark Energy Survey Collaboration} 2005, ArXiv: 0510346

\bibitem[\protect\citeauthoryear{{Vanzella}, {Cristiani}, {Fontana}, {Nonino},
  {Arnouts}, {Giallongo}, {Grazian}, {Fasano}, {Popesso}, {Saracco} \&
  {Zaggia}}{{Vanzella} et~al.}{2004}]{2004A&A...423..761V}
{Vanzella} E.,  {Cristiani} S.,  {Fontana} A.,  {Nonino} M.,  {Arnouts} S.,
  {Giallongo} E.,  {Grazian} A.,  {Fasano} G.,  {Popesso} P.,  {Saracco} P.,
  {Zaggia} S.,  2004, \aap, 423, 761

\bibitem[\protect\citeauthoryear{{Yeche}, {Petitjean}, {Rich}, {Aubourg},
  {Busca}, {Hamilton}, {Le Goff}, {Paris}, {Peirani}, {Pichon}, {Rollinde} \&
  {Vargas-Magana}}{{Yeche} et~al.}{2009}]{2009arXiv0910.3770Y}
{Yeche} C.,  {Petitjean} P.,  {Rich} J.,  {Aubourg} E.,  {Busca} N.,
  {Hamilton} J.~.,  {Le Goff} J.~.,  {Paris} I.,  {Peirani} S.,  {Pichon} C.,
  {Rollinde} E.,    {Vargas-Magana} M.,  2009, ArXiv: 0910.3770

\bibitem[\protect\citeauthoryear{{York} \& {SDSS Collaboration}}{{York} \&
  {SDSS Collaboration}}{2000}]{2000AJ....120.1579Y}
{York} D.~G.,  {SDSS Collaboration} 2000, \aj, 120, 1579

\end{thebibliography}
